\documentclass[%
 aip,
 amsmath,amssymb,
 reprint,%
]{revtex4-1}

\usepackage{graphicx}% Include figure files
\usepackage{dcolumn}% Align table columns on decimal point
\usepackage{bm}% bold math
\usepackage[utf8]{inputenc}
\usepackage[T1]{fontenc}
\usepackage{mathptmx}
\usepackage{etoolbox}

\makeatletter
\def\@email#1#2{%
 \endgroup
 \patchcmd{\titleblock@produce}
  {\frontmatter@RRAPformat}
  {\frontmatter@RRAPformat{\produce@RRAP{*#1\href{mailto:#2}{#2}}}\frontmatter@RRAPformat}
  {}{}
}%
\makeatother
\begin{document}

\preprint{AIP/123-QED}

\title{Capillary fluctuations and energy dynamics for flow in porous media}
% Force line breaks with \\
\author{James E. McClure}
  \email{mcclurej@vt.edu}
\affiliation{Virginia Polytechnic Institute \& State University, Blacksburg, Virginia}%Lines break automatically or can be forced with \\
\author{Steffen Berg}%
\affiliation{ 
Shell Global Solutions International B.V.
Grasweg 31,
1031HW Amsterdam,
The Netherlands
}%

\author{Ryan T. Armstrong}
\affiliation{%
University of New South Wales, Sydney
}%

\date{\today}% It is always \today, today,

\begin{abstract}
Capillary energy barriers have important consequences for immiscible fluid flow in porous media. 
We derive time-and-space averaging theory to account for non-equilibrium behavior and 
understand the role of athermal capillary fluctuations in the context of their relationship to larger scale phenomenological equations. The formulation resolves several key challenges associated with two-fluid flow in porous media:
(1) geometric and thermodynamic quantities are constructed as smooth functions of time based
on time-and space averages;
(2) averaged thermodynamics are developed for films;
(3) multi-scale fluctuation terms are identified, which account for transient behaviours of interfaces and films that occur due to pore-scale events;
(4) geometric constraints are derived and imposed on the averaged thermodynamics;
(5) a new constitutive model is proposed for capillary pressure dynamics that includes contributions from films;  and
(6) a time-and-space criterion for representative elementary volume (REV) is established based on capillary fluctuations.
Capillary fluctuations are assessed quantitatively based on 
pore-scale simulations and experimental core-flooding data.
\end{abstract}

\maketitle

\section{\label{sec:intro} Introduction}

The non-equilibrium response to fluid flow through porous materials
is inextricably linked to the length and time scales for processes 
that occur within the solid micro-structure \cite{Armstrong_Berg_2013,Berg_Ott_etal_13,Armstrong_Ott_etal_2014}. 
It is well known that fluid pressures fluctuate during immiscible displacement as a consequence
of pore-scale events \cite{Morrow_1970,Cueto-Felgueroso_Juanes_2015,Primkulov_2019}.
The associated fluctuations are athermal and cooperative in nature based on the fact
that they arise due to the influence of capillary forces 
\cite{Winkler_etal_2019}. Theoretical treatment of these fluctuations is of central importance
to modeling due to the fact that athermal fluctuations may not obey detailed
balance \cite{Gnesotto_2018}. Since the reversibility of molecular interactions is 
central to the development of Onsager's non-equilibrium theory, the validity of phenomenological
equations depends on the properties of fluctuations within the system \cite{Onsager_1931a,deGroot_Mazur_84}.
The validity of near-equilibrium approximations and Onsager's theory has been an important assumption in many theoretical developments for fluid flow in porous media \cite{Bear_Nitao_1995,marle1981multiphase,Hassanizadeh_Gray_93,Gray_Miller_14,Kjelstrup_2019}.

\begin{figure}[ht]
\centering
%\includegraphics[width=1.0\linewidth]{micro-emulsion-coalescence.png}
%\caption{Collision-coalescence mechanism within a micro-emulsion. 
%After two droplets collide, a bridge forms at the coalescence point.
%The larger droplet formed by the event evolves to attain
%a new equilibrium configuration based on the minimum potential energy.
%}
\includegraphics[width=1.0\linewidth]{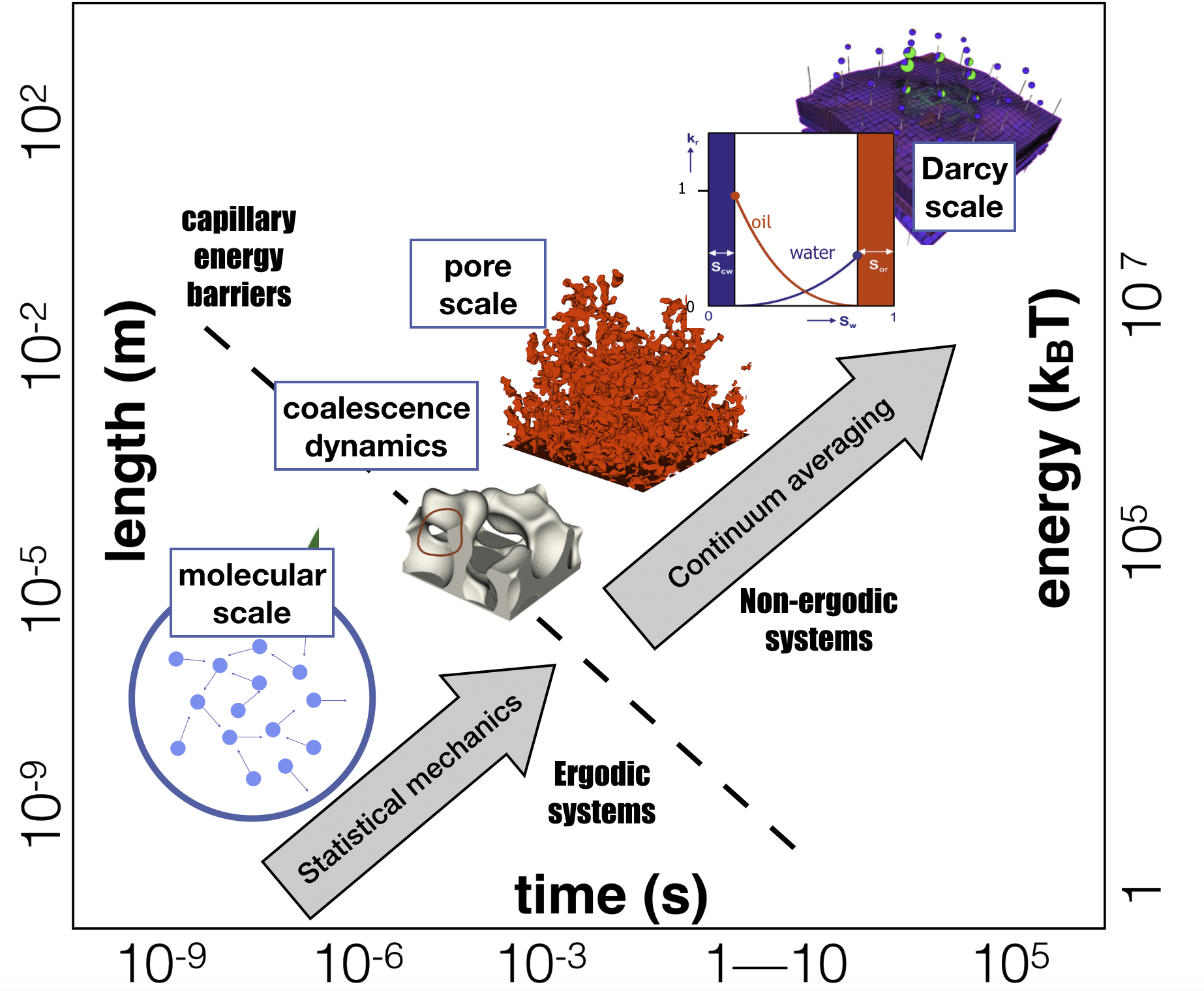}
\caption{Length and time scales for operative processes for 
multiphase flows in porous media. Geometric effects 
directly link smaller-scale phenomena with larger scale phenomena.
}
\label{fig:scales}
\end{figure}

When multiple length scales are present within a system, accompanying timescales 
will arise when considering the system dynamics. The linkage between temporal,
spatial and energy scales is summarized in Fig. \ref{fig:scales}.
Since information cannot propagate instantaneously, processes that occur over small length scales 
will tend to be more rapid compared to those that occur over larger length scales. 
Given a particular length scale, the energy scale relates to the work needed to
move matter and energy the requisite distance. The rate that information propagates in the system is embedded in the phenomenological coefficients associated with particular physical processes. With respect to the
validity of linear response theory, Onsager's theory holds remarkably well in small systems, implying that local near-equilibrium approximations are usually reasonable \cite{XU2006452}. 
At small length scales, near-equilibrium approximations are expected to fail only in 
extreme cases, such as in the immediate vicinity of shock waves \cite{Teller_1953}. However, as the length
scale associated with a process increases, so too does the required timescale to reach equilibrium.
For immiscible fluids in a geological setting, the timescale required for diffusive processes 
to reach equilibrium is remarkably slow; the timescale to reach equilibrium for
Ostwald ripening in CO$_2$-brine systems can be years \cite{Li_Garing_2020}. 
Within this context, slow physics represent 
a critical consideration when applying near-equilibrium assumptions at macroscopic length scales.
The slow relaxation of diffusive systems explains the near-ubiquitous failure of Fick's law in
porous media \cite{Cushman_Ginn_1993,Battiato_Tartakovsky_2011,Battiato_etal_2009,Cushman_OMalley_2015}. 
While the timescale associated with capillary fluctuations is measured in seconds, the timescale
to reach equilibrium can be hours to days in experimental systems \cite{Schluter_Berg_etal_2017}.
It is therefore of great interest to develop non-equilibrium theory that does not
rely on macroscopic near-equilibrium assumptions. In this paper we develop such a theory.

For multiphase flows in porous media, we identify three length scales of primary interest:
(1) the interfacial length scale associated with the width of the interface (10-50nm);
(2) the pore length scale associated with the solid micro-structure ($\sim$mm); and
(3) the Darcy length scale associated with macroscopic flow (cm - km).
For each length scale, particular processes of interest can be identified.
At the interface length scale, film swelling effects can occur due to flow;
coalescence events occur based on the merging of fluid regions, 
which form loops or otherwise alter the fluid topology \cite{McClure_etal_2020}.
Fluid singularities result from the associated disruption to the interface mean curvature
\cite{Pauslon_PRL_2008,Pak_etal_JPC_2018,Li_etal_nature_2018,Vahabi_etal_science_2018,Perumananth_etal_PRL_2019,Pauslon_PNAS_2012,Pauslon_NatureComm_2012,Dirk_2005,Ristenpart_2006,Wu_2004,Toro-Mendoza_2019,Orme_1997,Pauslon_PRL_2011,Pahlavan_2019}. 
Snap-off events are also common in typical flow processes \cite{Roof_1970}.
At the pore length scale, Haines jumps occur when fluid spontaneously fills regions of the the pore structure as 
fluids migrate between capillary energy barriers \cite{Haines_1930}.
These microscopic events are a primary source of non-equilibrium behavior, and are therefore a focus area when considering the physical meaning of near equilibrium approximations.
The length scale of interest for flow processes is the Darcy scale, typically orders of magnitude larger than that of the pore scale. 

In the following sections we provide a conceptual overview of the basic physical considerations
needed to model immiscible displacement in porous media. Formal theoretical development proceeds in six steps:
(1) definition of the thermodynamic system; 
(2) averaging in time-and-space;
(3) rate of external work in Darcy-scale systems;
(4) analysis for fluctuations in solid wetting energy; and
(5) derivation of geometric constraints;
(6) formulation of flux-force entropy and derivation of phenomenological equations for two-phase flow.
The derivation leads to explicit criteria for the validity of phenomenological equations 
based on the structure of capillary fluctuations in the system. Using results from simulation and
experimental data, we directly evaluate the derived capillary fluctuation terms and consider
the consequences for REV definition and the development of averaged descriptions for two-fluid
flow in porous media.

\section*{Background}

\begin{figure*}[ht]
\includegraphics[width=1.0\textwidth]{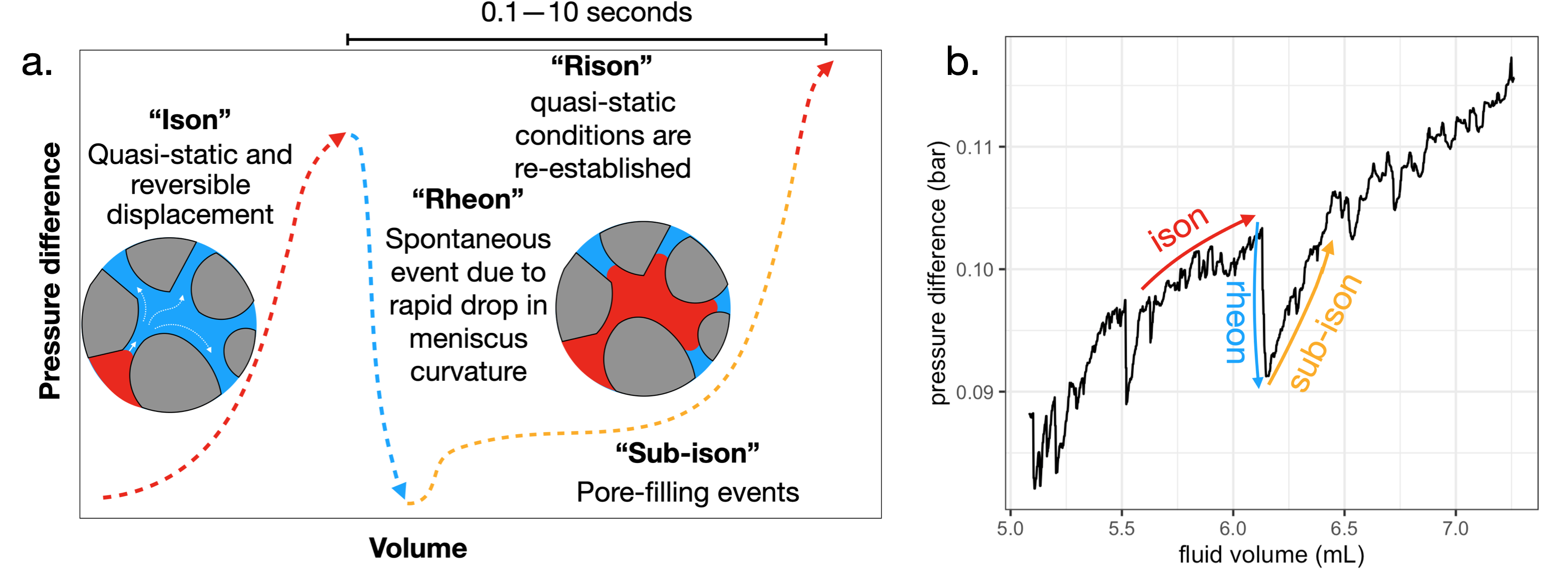}
\caption{Pore-scale displacement events can be separated into reversible events called
{\em isons} and irreversible {\em rheons} (a) Isons correspond to reversible displacement 
as the capillary pressure increases, forcing the fluid meniscus into narrower pore throats.
When the meniscus passes through the narrowest part of the pore-throat, energy is released
during the ensuing pore-filling event. A rapid drop in capillary pressure occurs as
the meniscus curvature decreases, which corresponds to a rheon. Fluid then spontaneously
fills the pore until the local capillary pressure is again balanced by the fluid pressure
difference; (b) identification of ison, rheon and sub-ison events from experimental data.}
\label{fig:rheon-ison}
\end{figure*}

At the pore scale, capillary forces are dominant and determine how fluids 
are distributed within the solid micro-structure. Capillary pressure accounts for the contribution of surface forces created by the surface energy and meniscus curvature \cite{Adamson_Gast_97}. The Young-Laplace equation relates the pressure difference between the adjoining fluids and the capillary pressure given by the product of the interfacial tension $\gamma$ and mean curvature,
\begin{equation}
     p_n - p_w = \gamma_{wn}\Big(\frac 1{R_1} + \frac 1{R_2} \Big) \;,
    \label{eq:laplace}
\end{equation}
where $R_1$ and $R_2$ are the principal curvature radii at points on the 
fluid meniscus. The Laplace equation holds only at mechanical equilibrium, 
and does not account for dynamic processes that involve moving interfaces. This
necessarily includes any process where the fluid volumes change, since these changes
can only occur based on movement of the interface. Situations also arise where interface 
movement occurs at constant volume, such as capillary waves. To describe these cases 
it is necessary to understand the non-equilibrium response of the fluid pressures based
on the capillary dynamics within the system.

As fluids squeeze through the microstructure of complex
materials subject to the influence of capillary forces, rapid changes to the fluid configuration
occur as the system jumps between metastable states. The displacement can be sub-divided
into a sequence of {\em isons} and {\em rheons} \cite{Morrow_1970,Cueto-Felgueroso_Juanes_2015,Berg_Slotte_2020}.
The associated events are depicted in Fig. \ref{fig:rheon-ison}. {\em Isons} correspond to reversible displacement events where the fluid meniscus invades a pore throat under quasi-static conditions, i.e. the fluid pressure difference is balanced by the capillary forces in the throat. The fluid pressure difference increases proportionate to the meniscus curvature as the interface is pushed into increasingly narrow parts of the pore throat. When the meniscus moves through the narrowest part of the pore neck, pore-filling occurs spontaneously based on an event often known as a Haines jump \cite{Haines_1930}. As the meniscus curvature decreases,
the fluid pressure difference significantly exceeds the capillary pressure based on
the interface curvature and fluid rapidly fills 
the pore. This irreversible event is known as a {\em rheon}. Pore-filling events that transpire as Haines jumps can dissipate a significant amount of energy \cite{Berg_Ott_etal_13,Armstrong_Berg_2013}. 
Furthermore, they occur very quickly (milliseconds to seconds) compared to typical laboratory- and 
reservoir-scale process (hours to weeks). Considering large systems, processes of interest involve very large 
numbers of such events. Understanding their overall effect is of central importance to understanding
transient effects for multiphase flows in porous media.

\begin{figure}[ht]
\centering
%\includegraphics[width=1.0\linewidth]{micro-emulsion-coalescence.png}
%\caption{Collision-coalescence mechanism within a micro-emulsion. 
%After two droplets collide, a bridge forms at the coalescence point.
%The larger droplet formed by the event evolves to attain
%a new equilibrium configuration based on the minimum potential energy.
%}
\includegraphics[width=1.0\linewidth]{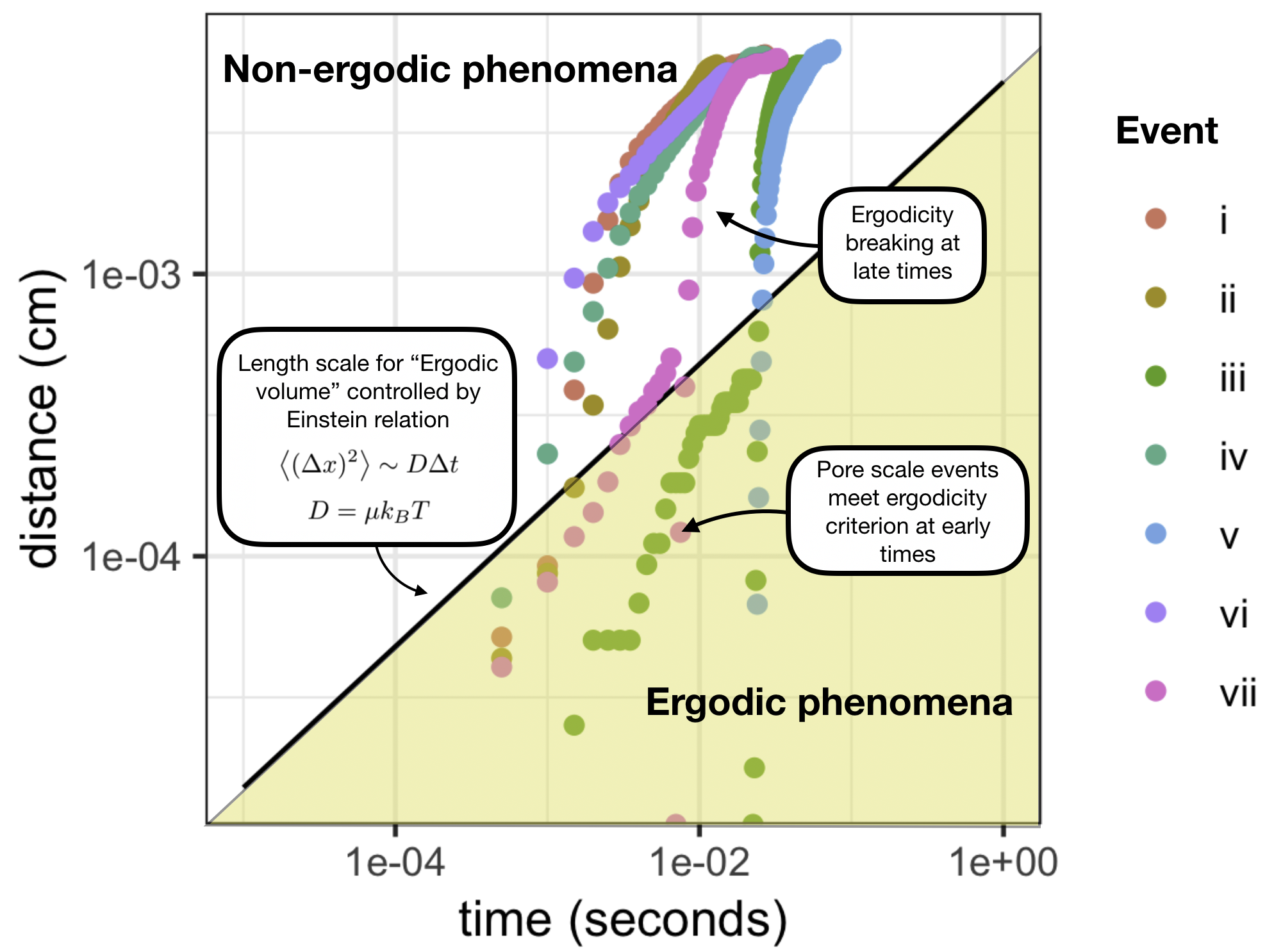}
\caption{The length scale for mixing can be estimated from the mean-squared distance for self-diffusion over a given length of time. Pore-scale events such as Haines jumps are 
associated with ergodicity breaking due to the rapid dynamics caused by the transfer of surface energy to other internal energy modes. Colored dots depict the observed trajectory
for seven different Haines jumps in a microfluidic system.
}
\label{fig:ergodicity}
\end{figure}

The basis to consider multiphase flow in porous media as a non-ergodic phenomenon 
is due to the rapid rate for local events at the pore-scale. In ergodic systems, 
energetic micro-states are considered to be locally well-mixed due to thermal effects.
Since the operative mixing mechanism is diffusive, the length scale for thermal mixing
can be estimated based on the self-diffusion coefficient and timescale \cite{McClure_etal_PRL_submitted}. 
When a mechanical event transfers energy more rapidly than diffusive thermal mechanisms, 
local symmetry breaking may occur. Fig. \ref{fig:ergodicity} considers the length and time scale for seven Haines jump events observed during displacement in a micro-fluidic system 
\cite{Armstrong_Berg_2013}. The distance traveled by the fluid meniscus 
is plotted versus the elapsed time for each event. This can be compared to the length scale for diffusive mixing during the elapsed time is $\Delta t$, given by 
\begin{eqnarray}
 \Delta x = \sqrt{D \Delta t}\;, 
 \label{eq:Einstein}
\end{eqnarray}
where $D=2.299\times10^{-5}$ cm$^2$ / sec  is the self-diffusion coefficient for water. We classify a phenomenon as ergodic
if the distance traveled over a particular time interval is smaller than the diffusive length scale. A phenomenon is classified as non-ergodic if the distance traveled during an interval of time exceeds this distance. 
Fig. \ref{fig:ergodicity} clearly shows that Haines jumps begin in the ergodic region but 
cross into the non-ergodic regime as capillary energy is converted into kinetic energy. 
%Since the energy for the diffusive modes is much smaller, $\sim k_B T$, the thermal energy is 
Over the timescale for a Haines jump, the length scale for thermal mixing is an order of
magnitude smaller than the characteristic length scale for the event. The system can therefore be considered to be thermally well-mixed at the scale of $\sqrt{D \Delta t}$, but not at the 
scale of a Haines jump.

At the Darcy scale, energy that is dissipated from pore-scale events can be estimated from basic concepts of work and energy.  A typical experiment can be treated as a system where 
external work is performed
in two primary ways. First is the work associated with fluid flow. The volumetric flow rate is ${\mathbf{Q}}_i$
for each fluid $i \in \{w,n\}$ results due to external work performed on the system
by some external potential $\Phi_i$, e.g. due to an
external pressure gradient or body force. Since work is defined as the applied force multiplied
by the distance, the total rate of work is
\begin{equation}
   \frac{d W_{darcy}}{dt}  =   \sum_{i\in\{w,n\}} \Delta \Phi_i {\mathbf{Q}}_i \cdot \mathbf{n}_b  \;,
   \label{eq:work-darcy}
\end{equation}
where $\Delta \Phi_i  = \Phi_i^{(in)} - \Phi_i^{(out)}$ is the potential 
difference between the inlet and the outlet and $\mathbf{n}_b$ is the boundary
normal vector. This expression is is directly implied from the definition of mechanical work. 
The boundary terms can also be obtained by applying the divergence theorem in consideration of
the internal stresses \cite{Berg_Slotte_2020}.
Conceptually, the work due to Darcy flow occurs in a steady-state system, meaning that
there are no net changes to the internal energy, with each fluid forming static connected pathways through the sample that allow flow to occur. The conventional relative permeability relationship predicts the flux per area as
\begin{equation}
{\mathbf{q}}_i =  - \frac{k_{ri}}{\mu_i} \mathbf{K} \cdot \nabla {\Phi}_i  \;,
\label{eq:kr-sw}
\end{equation}
where  ${\mathbf{q}}_i = {\mathbf{Q}}_i /A$, $k_{ri}$ is the relative permeability, 
$\mu_i$ is the dynamic viscosity and
$\mathbf{K}$ is the permeability tensor for the material. This can be combined with 
Eq. \ref{eq:work-darcy} to approximate the rate of work based on an applied external 
potential gradient. The rate of energy dissipated within the system is thereby
embedded in $k_{ri}$.

Pressure-volume work occurs when one fluid displaces another. Considering a wetting and 
non-wetting fluid with pressure $p_w$ and $p_n$, this contribution to the external work is given by
\begin{equation}
    \frac{\partial  W_{ext}}{\partial t} =   (p_w - p_n) \frac{\partial V_w}{\partial t} \;.
    \label{eq:work-ext}
\end{equation}
External work performed on the system is transferred to the internal energy modes, 
which include contributions to the internal thermal energy from
dissipated heat. Changes to the surface energy account for a significant part of the 
external pressure-volume work. If the solid material is water-wet, 
the change in surface energy is proportional to the change in surface
area based on the interfacial tension $\gamma_{wn}$. Morrow therefore defined the displacement efficiency as the fraction of pressure-volume
work that is converted to surface energy \cite{Morrow_1970}. Seth and Morrow showed that the efficiency for a primary drainage varies from 10--95\% depending on the material type \cite{Seth_Morrow_2007}. The dissipated energy can also be estimated from the 
mechanistic picture illustrated in Fig. \ref{fig:rheon-ison}. The pressure-volume work
associated with an initial {\em ison} can be considered to be reversible based on the fact
that the displacement is quasi-static. Since a {\em rheon} is spontaneous, the work 
associated with the ensuing sub-ison accounts for the dissipated energy after any 
surface energy changes generated by the event has been subtracted. With energy
input into the system given by Eqs. \ref{eq:kr-sw} and \ref{eq:work-ext}, the
internal energy dynamics of the system can be treated based on conservation of energy. 
%Comparing capillary pressure curves to the effective permeability for typical materials,
%the rate of work for displacement is two orders of magnitude larger
%than the rate of work for connected pathway flow {\color{red} @Steffen -- Reference to cite here?}. 

Models for capillary pressure dynamics have been previously developed 
to account for the macroscopic behaviour during displacement. The first such models
were developed using concepts of work and energy \cite{Morrow_1970}. Subsequent models
incorporated concepts of non-equilibrium thermodynamics \cite{marle1981multiphase,Hassanizadeh_Gray_93} 
and kinematics \cite{Gray_Miller_14,Gray_Dye_etal_15}. Previously developed models fail to  deal with the challenges associated with the non-smooth response of the fluid pressure and interfacial curvature. This critical challenge is associated with changes to the fluid topology, which necessarily cause a discontinuity in the interfacial curvature \cite{McClure_etal_2020,Pauslon_NatureComm_2012,Pahlavan_2019}. 
The apparent geometric discontinuity links to questions about whether or not 
multi-fluid systems can be considered to be in a near-equilibrium state.
Bear \& Nitao argue that while relatively simple macroscopic thermodynamic equations of state must exist at equilibrium due to the Gibbs phase rule, these relations will
break down under non-equilibrium conditions if the mean-square deviation is
too large \cite{Bear_Nitao_1995}. For systems with slow physics, near equilibrium conditions can break down as the length scale for the REV increases due to the 
larger timescale required for mixing to occur. 
We argue that the breakdown of near-equilibrium behavior
will occur at large length scales based on the argument summarized in Fig. \ref{fig:ergodicity}.
Near-equilibrium conditions should hold at spatial scales smaller than $\sqrt{D\Delta t}$,
with $\Delta t$ being the timescale for pore-scale events. Averaging in time-and-space
provides a way to upscale the formulation to larger length and time scales based on 
a near-equilibrium approximation at a small scale. Upscaling is then performed based on integrals,
which do not require any near-equilibrium approximation at at the larger scale. In this work, we consider how multi-scale effects influence the transient  capillary pressure behavior and consider consequences for the rate of energy dissipation. Multi-scale fluctuation terms are identified explicitly, which can be directly evaluated and compared to other energy
dynamics within the system. Conditions derived based on these fluctuation terms clearly 
demonstrate when Onsager theory can be applied to derive phenomenological models such as the conventional relative permeability relationship. A new expression for the dynamic capillary pressure is also derived, which directly includes effects due to fluctuations. Capillary fluctuations are then directly evaluated based on pore-scale simulations and experimental data.

\section{Theoretical Approach}

In the following sections we present the theoretical approach used to derive constitutive models
for immiscible fluid flow through porous media. Our approach relies on averaging in both time and space,
which is distinct from previously developed models. Capillary fluctuation terms are included
explicitly, and special treatment is needed to develop a flux-force entropy inequality that accounts for the associated energy dynamics. 
The flux-force form of the entropy inequality can be exploited to 
derive standard expressions for relative permeability as well as a dynamic capillary pressure relationship to describe unsteady displacement. The theoretical approach is organized into six steps:
\begin{enumerate}
    \item {\bf Thermodynamic description} -- variables required to formulate a classical thermodynamic description 
    at a scale where the behavior is ergodic, including effects due to fluid phases, interfaces and films;
    \item {\bf Averages in space and time} -- averaging is applied to develop thermodynamic forms
    that hold at larger length and time scales. Fluctuation terms arise as a consequence of averaging in both space
    and time.
    \item {\bf Rate of work and entropy inequality} -- an entropy inequality is developed
    by defining the rate of external work associated with fluid flow; 
    \item {\bf Surface wetting fluctuations} -- analysis is performed to interpret the
    surface wetting fluctuation term; 
    \item {\bf Geometric constraints} -- a geometric constraint is needed to relate the time derivative for the interfacial area to the time derivative for the fluid volume fractional. 
    A relationship between partial derivatives for geometric quantities is derived;
    \item {\bf Flux-force form of entropy inequality} -- all results are combined to form a flux-force
    form of the entropy inequality. A REV constraint is derived for the fluctuations. Provided the constraint is
    met, standard relative permeability relationships are obtained as well as a form to approximate the 
    capillary pressure dynamics during unsteady displacement.
\end{enumerate}
By explicitly considering the contribution from fluctuations in the flux-force form of the entropy inequality,
conditions for the validity of Darcy-scale phenomenological equations can be developed explicitly. The derived condition is a constraint on the representative elementary volume (REV) for the system, which considers both
spatial and temporal aspects. The theory does not rely on the assumption of detailed balance at the macroscopic scale, as ergodic conditions are
assumed to hold only at small length scales.
The derived REV condition is a constraint that fluctuations must obey to derive phenomenological equations that 
provide a valid representation for the entropy production in the system. 
Experimental data is then used to assess the role of capillary fluctuations in the context of the theory.

\subsection{Thermodynamic Description}

Thermodynamics provides a mechanism to link flow processes with conservation of energy. For non-equilibrium systems, this is accomplished by 
deriving an expression for the entropy production, which can then be exploited to derive phenomenological equations for particular processes \cite{deGroot_Mazur_84}. At the practical
level, thermodynamics is a multi-scale book-keeping system to keep track of how energy is distributed within a system. 
%Entropy production occurs when internal energy is transferred to thermal modes that are linked to the system entropy based on the molecular degrees of freedom. 
The thermodynamics of heterogeneous systems are typically treated by dividing the system into sub-regions based on the system composition 
 \cite{Kjelstrup_Bedeaux_08,Kjelstrup_2019,Galteland_2019,Kjelstrup_2019,Galteland_2019}. 
In addition to the entropy $S$ we consider the extensive variables to be
\begin{itemize}
    \item $V_i$ -- fluid volume  for $i\in \{w, n\}$
    \item $A_{n}$ -- surface area for fluid $n$
    \item $A_{s}$ -- surface area for the solid material
    \item $h_s$ -- film thickness along the solid material
    \item $N_{ik}$ -- number of molecules of type $k$ within any region $i \in \{w, n,s\}$
\end{itemize}
For simplicity energy associated with the solid is omitted, such as
effects associated with deformation. Considering solid that does not 
deform, the surface area bounding the wetting fluid $A_w$ is omitted based on the fact that only two of $\{A_n,A_w,A_s\}$ are independent. The internal energy is given as a function of the independent extensive quantities
\begin{equation}
    U = U(S,V_w,V_n,A_{n},A_s,h_s,N_{ak},N_{bk},N_{sk})\;.
    \label{eq:extensive-dependence}
\end{equation}
 The Euler equation for the internal energy of the system is then written as
 \begin{eqnarray}
    U &=& T S - p_w V_w -p_n V_n + \gamma_{wn} A_{n}  + \gamma_s A_s - \Pi_s h_s A_s \nonumber \\
    && + \mu_{ak} N_{ak} + \mu_{bk} N_{bk} + \mu_{sk} N_{sk} \;,
    \label{eq:euler}
\end{eqnarray}
where the intensive measures are
\begin{itemize}
    \item $p_i$ -- fluid pressure for $i\in \{w, n\}$
    \item $\gamma_{wn}$ -- surface energy for fluid-fluid interface
    \item $\gamma_{s}$ -- surface energy for the solid material
    \item $\Pi_s$ -- disjoining pressure along the solid material
    \item $\mu_{ik}$ -- chemical potential for molecule $k$ within any region $i \in \{w, n,s\}$
\end{itemize}
In a strongly wetted system, the solid surface is coated everywhere by a thin film of fluid $w$.
This means that fluid $n$ interacts with $w$ everywhere along its boundary, described by the surface
energy $\gamma_{wn}$. If the system is not strongly wetted by $w$, the formulation can still be used
based on the fact that any excess energy due to interactions between the fluids and the solid 
will be accounted for by local variations in $\gamma_{s}$ along the solid. The complexity of the 
fluid-solid interactions is therefore embedded within the associated energy term, $\gamma_s A_s$.
In our approach, we do not treat the solid as being in contact with a particular fluid. Instead we consider the situation where the associated interfacial energy $\gamma_s$ and film thickness
$h_s$ can vary along the solid. The formulation presupposes that films coat the solid surface everywhere, with the understanding that if the film thickness $h_s=0$ locally on portions of the solid 
surface, this will be equivalent to treating films as covering only part of the solid surface. 
The disjoining pressure is defined as 
\begin{equation}
    \Pi_s \equiv - \frac{1}{A_s} \Big(\frac{\partial U}{\partial h_s}\Big)_{S,V_w,V_n,A_n,A_s,N_{ak},N_{bk},N_{sk}}
\end{equation}
The energy due to fluid-solid interactions is therefore lumped into two terms, one associated with the
interfacial energy and a second due to the disjoining pressure and film contributions.
Variations in $\gamma_s$ and $h_s$ along the surface of the solid
occur due to the interaction between the fluid and solid material, including
the effects of roughness and chemical heterogeneity. 

The terms that account for capillary fluctuations are neglected in
conventional non-equilibrium thermodynamics; these appear only when time-averages are introduced. When considering flow processes that occur over long periods of time, it is easy to overlook fast 
dynamics. Haines jumps are one such category of event, with dynamics that typically occur over a duration of
milliseconds to seconds \cite{Mohanty_etal_1987,DiCarlo_2003,Berg_Ott_etal_13}. 
Topological changes such as droplet coalescence also dissipate energy, occurring even 
more rapidly and contributing to pressure fluctuations. While both classes of event are common for flows in porous media, the associated energy dissipation has not been properly included in thermodynamic models. In the following sections, we demonstrate that pressure fluctuations represent an essential contribution to the energy dynamics for two-fluid flow in porous media. 

\subsection{Thermodynamic Averages in Space and Time}
\label{sec:equilibrium}

The use of time averaging to study non-equilibrium behavior for two-fluid flow
in porous media was first treated by Aryana and Kovscek
\cite{Aryana_2013}. While these authors demonstrated that the conventional
Buckley-Leverett model could be derived based on this approach (as well
as extended models), aspects pertaining to the non-equlibrium thermodynamics were not considered. Here we develop this theory explicitly.
Averages are constructed with the
time-and-space averaging operator, defined as 
\begin{equation}
    \big< f \big> \equiv\frac{1}{\lambda \mathcal{V}} \int_{\Lambda} \int_{\Omega} f dV dt\;,
    \label{eq:average}
\end{equation}
where $\Lambda$ is a region of time with duration $\lambda$ and $\Omega$ is a spatial averaging
region \cite{McClure_etal_PRL_submitted}. $\mathcal{V}$ is the volume associated with thermodynamic measurements
which can be estimated from the length scale given in Eq. \ref{eq:Einstein}. 
For example, if the fluid pressure is measured with a transducer, $\mathcal{V}$ can be thought of as
the volume of fluid that is in local equilibrium with the measured value as a consequence of 
local mixing due to motion of molecules in the immediate vicinity. 
The total volume of the spatial averaging region $\Omega$ is $V > \mathcal{V}$. 
Dividing Eq. \ref{eq:euler} by $\mathcal{V}$ leads to a representation of the internal energy
on a per-unit-volume basis. Since a point in space may be within one fluid or the other, we define
the indicator function
\begin{equation}
    \phi_i (\mathbf{x}) = \left \{  \begin{array}{cc} 1& \mbox{if $\mathbf{x} \in \Omega_i$} \\
    0 & \mbox{otherwise} \end{array}\right.
    \label{eq:subset}
\end{equation}
where $i \in \{w,n\}$ to denote the region of space occupied by each fluid,
$\Omega_i$. Integrating $\phi_i$ over any region will therefore return the 
volume of the associated fluid within that region. 

Statistical definitions for thermodynamic quantities rely on the ergodic hypothesis such that ensemble averages are the same as averages in space or time. The basis for this assumption is that the energy micro-states are locally well-mixed, i.e. that the molecules in the system interact and exchange energy on a timescale that is sufficiently fast relative to the size of the system considered. Within this context Eq. \ref{eq:euler} holds in
the limit of infinite time and non-equilibrium expressions must be developed to study
time-dependent processes. In the context of ergodic theory, the averaging operator given by Eq. \ref{eq:average} explicitly mixes information in space and time to generate averages. 
Non-ergodic effects can be accounted for in the averaged system provided that 
$\Lambda$ and $\Omega$ are chosen appropriately. Averages of the extensive quantities are 
simple averages so that these quantities retain their basic meaning
\begin{eqnarray}
    &\overline{S} \equiv \big< S \big>\;, \quad \overline{V}_i \equiv  V\overline{\phi}_i = V\big< \phi_i \big>\;, \quad 
    \quad  \overline{N}_{ik} \equiv \big< N_{ik} \big>\;, &  \nonumber \\
    &\overline{A}_i \equiv \big< A_i \big>\;,  \quad \overline{h}_s \equiv \big< h_s \big> \;, &
    \label{eq:extensive}
\end{eqnarray}
The intensive measures are then defined to maintain scale-consistency with Eq. \ref{eq:euler}
\begin{equation}
      \overline{T} \equiv \frac{\big< T S \big>}{\big< S \big>}\;, \quad \overline{p}_i \equiv \frac{\big< p_i \phi_i \big>}{\big< \phi_i \big>}\;, \quad
    \overline{\mu}_{ik} \equiv \frac{\big< \mu_{ik} N_{ik} \big>}{\big< N_{ik} \big>}\;,  \nonumber
\end{equation}
\begin{equation}
    \overline{\gamma}_i \equiv \frac{\big< \gamma_i A_i \big>}{\big< A_i \big>}\;, \quad  
    \overline{\Pi}_s \equiv \frac{\big< \Pi_s A_s h_s \big>}{\big< A_s h_s \big>}\;.
    \label{eq:intensive}
\end{equation}
The internal energy in the averaged thermodynamic model will then be given by
 \begin{eqnarray}
    \overline{U} &=& \overline{T} \overline{S} - \overline{p}_w \overline{V}_w -\overline{p}_n \overline{V}_n + \overline{\gamma}_{wn} \overline{A}_{n}  + \overline{\gamma}_s \overline{A}_s - \overline{\Pi}_s \overline{A}_s \overline{h}_s \nonumber \\
    && + \overline{\mu}_{ak} \overline{N}_{ak} + \overline{\mu}_{bk} \overline{N}_{bk} + \overline{\mu}_{sk} \overline{N}_{sk} \;.
\end{eqnarray}

In a non-equilibrium system, deviations of the intensive quantities contribute to the energy dynamics
\cite{McClure_etal_PRL_submitted}. 
The deviation terms are defined as the difference between the microscopic value and the averaged value
\begin{eqnarray}
   &T^\prime \equiv T - \overline{T} \;, \quad
   p_i^\prime\equiv p_i - \overline{p}_i \;, \quad
   \mu_{ik}^\prime\equiv \mu_{ik} - \overline{\mu}_{ik} \;, & \nonumber \\
 & \gamma_i^\prime\equiv \gamma_i - \overline{\gamma}_i \;, \quad
   \Pi_s^\prime\equiv \Pi_s - \overline{\Pi}_s \;. &
\end{eqnarray}
The time derivative for the internal energy includes contributions from the time derivatives of the averages of the extensive
measures as well as fluctuation terms that arise due to the microscopic deviations in the intensive quantities
 \begin{eqnarray}
    \frac {\partial \overline{U}}{\partial t} &=& 
    \overline{T} \frac {\partial \overline{S}}{\partial t}
    - \overline{p}_w \frac {\partial \overline{V}_w }{\partial t}
    -\overline{p}_n \frac {\partial \overline{V}_n}{\partial t} 
    + \overline{\mu}_{ik} \frac {\partial \overline{N}_{ik} }{\partial t}
    + \overline{\gamma}_{wn} \frac {\partial \overline{A}_{n}  }{\partial t}
    \nonumber \\ 
&&
%    +\overline{\Pi}_{wn} \frac {\partial \overline{H}_{wn} }{\partial t}
    + \big(\overline{\gamma}_s - \overline{\Pi}_s\overline{h}_s\big) \frac {\partial \overline{A}_s }{\partial t}
    - \overline{\Pi}_s \overline{A}_s \frac {\partial \overline{h}_s}{\partial t} 
    -\Big< S \frac{\partial{T}^\prime}{\partial t}\Big> \nonumber \\ 
&&  
    + \Big< V_w \frac{\partial{p}_w^\prime}{\partial t}\Big>
    + \Big< V_n \frac{\partial{p}_n^\prime}{\partial t}\Big> 
    - \Big< N_{ik} \frac{\partial{\mu}_{ik}^\prime}{\partial t}\Big>\ 
    \nonumber \\ 
&&  
    - \Big< A_{n} \frac{\partial{\gamma}_{wn}^\prime}{\partial t}\Big>
    - \Big< A_s \frac{\partial{\gamma}_s^\prime}{\partial t}\Big>
%    + \Big< h_f \frac{\partial{\Pi}_f^\prime}{\partial t}\Big>
    + \Big< A_s h_s \frac{\partial{\Pi}_s^\prime}{\partial t}\Big>  \;,
    \label{eq:non-eq-thermo}
\end{eqnarray}
where a sum is implied for all chemical potential terms. The multi-scale fluctuation terms account for transient effects and heterogeneity associated with the thermodynamic behaviour:
\begin{itemize}
    \item $\Big< V_i \frac{\partial{p}_i^\prime}{\partial t}\Big>$ -- fluid pressure fluctuations, e.g. due to the transient effect of changing capillary forces on the fluid pressure during flow;
    \item $\Big< A_{n} \frac{\partial{\gamma}_{wn}^\prime}{\partial t}\Big>$ -- fluctuations to the fluid-fluid interfacial tension, e.g. due to different surface tension along the fluid-fluid interface, which may arise if surfactants are present;
    \item $\Big< A_s \frac{\partial{\gamma}_s^\prime}{\partial t}\Big> $ -- fluctuations to the fluid-solid surface energy, e.g. which occur due to movement of the contact line and complex fluid-solid interactions that occur due to surface heterogeneity;
    \item $\Big< A_s h_s \frac{\partial{\Pi}_s^\prime}{\partial t}\Big> $ -- fluctuations to the disjoining pressure of films on the solid surface, e.g. which can occur due to localized film swelling; 
    \item $\Big< N_{ik} \frac{\partial{\mu}_{ik}^\prime}{\partial t}\Big>$ -- fluctuations to chemical potential, which 
    arise due to mass diffusion.
\end{itemize}

There is a direct link between the pressure fluctuation defined in Eq. \ref{eq:non-eq-thermo},
and rugged energy landscapes depicted in Fig. \ref{fig:rheon-ison}. 
Pressure fluctuation terms contribute to the non-equilibrium description
based on pore-scale events that transpire within the averaging window defined by
$\Lambda$ and $\Omega$. The reversible work associated with an {\em ison} is captured
based on the product of the fluid pressure difference and rate of volume change, as in standard
non-equilibrium thermodynamics. During a strictly quasi-static displacement, the rate of change for the 
microscopic pressure $p_i$ should closely correspond with the average fluid pressure $\overline{p}_i$.
On the other hand, if we consider a {\em rheon} to be a rapid pressure change at 
constant volume, then $p_i$ must deviate significantly from $\overline{p}_i$ over the duration
of the event. 
%The {\em rheon} includes no pressure-volume work, since $\partial V_n / \partial t =0$, and the associated energy change is described by the fluctuation terms. 
Therefore, if there are {\em rheons} occuring within $\Omega$ during the time interval
$\Lambda$ then the associated fluctuation terms are expected to have a significant effect. 
 This will be treated in more detail in sections to follow.

\subsection{Rate of work and entropy inequality}

The change to the internal energy of the system is determined based on the rate of work
done on the system and the heat added
\begin{equation}
     \frac{\partial \overline{U}}{\partial t} =  \frac{\partial \overline{W}}{\partial t} +  \frac{\partial \overline{Q}}{\partial t}\;.
     \label{eq:energy-work-heat}
\end{equation}
The instantaneous rate of work is given by the sum of Eqs. \ref{eq:work-darcy}
and \ref{eq:work-ext}, the latter being already built into the thermodynamics.
For the data considered in this work, which correspond to immiscible two-fluid flow in strongly water-wet systems, several assumptions can be made to simplify the form of Eq. \ref{eq:non-eq-thermo}:
\begin{enumerate}
    \item the system temperature is uniform and no heat is added to the system, which is justified by the large heat capacity for the system,
        \[ T = \overline{T} \rightarrow \Big< S \frac{\partial T^\prime}{\partial t}\Big> = 0\;,
        \quad \frac{\partial Q}{\partial t} = 0\; ; \] 
    \item the fluid-fluid interfacial tension is constant everywhere:
        \[ \gamma_{wn} = \overline{\gamma}_{wn} \rightarrow \Big< A_{n} \frac{\partial{\gamma}_{wn}^\prime}{\partial t}\Big> = 0 \;;
        \]
    \item the solid does not change shape, meaning that 
    \[ 
       \frac{\partial{\overline{A}_s}}{\partial t} = 0 \;, \quad
       \frac {\partial \overline{V}_w }{\partial t} = - \frac {\partial \overline{V}_n }{\partial t}\; ;
    \]
    \item compositional effects can be ignored 
    \[ \overline{\mu}_{ik} \frac {\partial \overline{N}_{ik} }{\partial t}=0 \;,
        \quad \Big< N_{ik} \frac{\partial{\mu}_{ik}^\prime}{\partial t}\Big> =0\; .\]
\end{enumerate}
Subject to these assumptions and using Eq. \ref{eq:energy-work-heat} to replace the change to the
internal energy with the rate of external work done on the system,
Eq. \ref{eq:non-eq-thermo} can be simplified and rearranged to obtain an entropy inequality
 \begin{eqnarray}
       \frac {\partial \overline{S}}{\partial t}
        &=& 
\frac{1}{\overline{T} } \Bigg \{
   \underbrace{
   L\Big[\nabla \overline{\Phi}_w \cdot \overline{\mathbf{Q}}_w
    + \nabla \overline{\Phi}_n \cdot \overline{\mathbf{Q}}_n \Big]}_{\mbox{work of fluid flow}}
    \nonumber \\ 
&&
    +    V \Big[ 
    \underbrace{(\overline{p}_w -\overline{p}_n ) \frac {\partial \overline{\phi}_w }{\partial t}}_{\mbox{ison}}
    - \underbrace{\Big< \phi_w \frac{\partial{p}_w^\prime}{\partial t} + 
    \phi_n \frac{\partial{p}_n^\prime}{\partial t}\Big>}_{\mbox{rheon}}\Big]
  \nonumber \\ 
&&  
    - \underbrace{\overline{\gamma}_{wn} \frac {\partial \overline{A}_{n}  }{\partial t}
    + \Big< A_s \frac{\partial{\gamma}_s^\prime}{\partial t}\Big>}_{\mbox{change in surface energy}}
  \nonumber \\ 
&&  
    + \underbrace{\overline{A}_s \overline{\Pi}_s \frac {\partial \overline{h}_s}{\partial t} 
    - \Big< A_s h_s \frac{\partial{\Pi}_s^\prime}{\partial t}\Big>}_{\mbox{film swelling}} \Bigg\} \ge 0 \;,
    \label{eq:entropy-inequality}
\end{eqnarray}
where $L$ is the distance between the sample inlet and outlet
and the time-average for the  fluid volume fraction is
\begin{equation}
%    \phi_i = \frac{V_i}{V}\;, \quad 
    \overline{\phi}_i = \frac{\overline{V}_i}{V} 
    \quad  \mbox{for $i \in \{w, n\}$} \;.
\end{equation}
%The surface area per unit volume and the associated average are
%\begin{equation}
%    a_i = \frac{A_i}{V}\;, \quad \overline{a}_i = \frac{\overline{A}_i}{V} 
%    \quad  \mbox{for $i \in \{n,s\}$} \;.
%\end{equation}
%and the film thickness per unit volume and the associated average are
%\begin{equation}
%    h_s = \frac{H_s}{V}\;, \quad \overline{h}_s = \frac{\overline{H}_s}{V} \;.
%\end{equation}
In the context of Fig. \ref{fig:rheon-ison}, specific terms appear that correspond with 
the reversible pressure-volume work associated with an {\em ison} and the spontaneous {\em rheons}. Note that the average pressure will behave as a smooth function of time, meaning that the ``rugged" parts of the energy landscape can be fully transferred to the fluctuation terms if $\Lambda$
is large enough. Three fluctuation terms are retained, the first associated with capillary fluctuations in the
fluid pressures. The second is due to wetting fluctuations that are attributed to the movement of the
contact line during fluid displacement, which appears due to the time derivative for the deviation 
of the solid interaction energy $\gamma^\prime_s$. Intuitively, the fluid-solid interaction
energy changes when the contact line moves. Finally, the effects of film swelling 
contribute based on the fluctuation of the disjoining pressure, which will change based on changes to 
the film thickness. This contribution is important
primarily when the wetting fluid is poorly connected, in which case films that coat the solid surface 
provide a primary flow pathway for that fluid. To obtain a more useful form of the entropy inequality,
a flux-force form must be obtained \cite{Onsager_1931a}. In the following sections we demonstrate how 
this can be accomplished for systems where the fluctuation terms play a significant role.

\subsection{Fluctuations and solid wetting energy}

The model considered here is developed for strongly water-wet systems where the solid
is coated everywhere by a thin water film. This means that the boundary of the non-wetting
fluid touches water everywhere, and the surface energy is proportional to the total
boundary surface area $A_n$. This simplifies the geometric analysis
presented in the next section.
For the situations considered here, where the surface wetting properties are homogeneous
and do not change with time, the surface wetting fluctuation is zero,
\begin{equation}
\frac{\partial{\gamma}_s^\prime}{\partial t} =  0 \rightarrow \Big< A_s \frac{\partial{\gamma}_s^\prime}{\partial t}\Big> =0
\end{equation}
In many situations the surface wetting properties will be heterogeneous, with the local
wetting energy varying based on chemical properties and roughness of the solid.
In these more general conditions, all of the the solid surface energy contributions 
are embedded into a single term. This term can also account for situations where the solid is not strongly wetted by one fluid. In this case only part of the solid is in contact with water, with associated surface area $A_{ws}$. The rest of the solid contacts the non-wetting fluid, with $A_{ns} = A_{s} + A_{ws}$. In this situation the fluctuation term can be subdivided into distinct regions of water and non-wetting fluid contact the grain surface. 
\begin{equation}
\Big< A_s \frac{\partial{\gamma}_s^\prime}{\partial t}\Big>
=     \Big< \overline{A}_{ns} \frac{\partial{\gamma}_s^\prime}{\partial t} \Big>+
   \Big<  \overline{A}_{ws} \frac{\partial{\gamma}_s^\prime}{\partial t} \Big> = 0
   \label{eq:wetting-i}
\end{equation}
If the solid surface energy is constant over each region with $\gamma_s = \overline{\gamma}_{ws}$
everywhere on $A_{ws}$  and $\gamma_s = \overline{\gamma}_{ns}$ everywhere on $A_{ns}$, then
\begin{equation}
\overline{\gamma}_s = \frac{\big<\gamma_s A_s \big>}{\overline{A}_s}
=  \frac{\overline{\gamma}_{ns}\overline{A}_{ns} + \overline{\gamma}_{ws} \overline{A}_{ws}}{\overline{A}_s}\;.
\label{eq:wetting-ii}
\end{equation}
In this case  fluctuations of the solid surface energy are determined by movement of the contact line, which changes the surface areas. Since $\overline{\gamma}_{ns}$, $\overline{\gamma}_{ws}$
and $\overline{A}_s$ are constant with respect to time,
\begin{equation}
%    \frac{\partial (\overline{\gamma}_{ns} - \overline{\gamma}_{s})}{\partial t}
\frac{\partial  \overline{\gamma}_{s}}{\partial t}
=  \frac{1}{\overline{A}_s} \Big( \overline{\gamma}_{ns} \frac{\partial \overline{A}_{ns} }{\partial t}
+\overline{\gamma}_{ws} \frac{\partial \overline{A}_{ws} }{\partial t}\Big)\;.
\label{eq:wetting-iii}
\end{equation}
Since the total solid surface area is constant
\begin{equation}
\frac{\partial \overline{A}_{ns} }{\partial t}= -\frac{\partial \overline{A}_{ws} }{\partial t} \;.
\label{eq:wetting-iv}
\end{equation}
Using this result with Eqs. \ref{eq:wetting-ii} and \ref{eq:wetting-iii} and doing some basic 
algebra shows that the fluctuation term is 
\begin{eqnarray}
\Big< A_s \frac{\partial{\gamma}_s^\prime}{\partial t}\Big>
&=&     
\Big(  \frac{\overline{A}_{ns}+ \overline{A}_{ws}}{\overline{A}_s} \Big)
 \Big( \overline{\gamma}_{ws}- \overline{\gamma}_{ns}\Big)
 \frac{\partial \overline{A}_{ns} }{\partial t}  \nonumber \\
 &=&     
 \Big( \overline{\gamma}_{ws}- \overline{\gamma}_{ns}\Big)
 \frac{\partial \overline{A}_{ns} }{\partial t} \;.
 \label{eq:fluctuation-wetting-v}
\end{eqnarray}
This is the conventional expression for change in fluid-solid surface energy.
Note that while $A_{ns}$ and $A_{ws}$ are extensive quantities, only their sum, $A_s$ was included in Eq. \ref{eq:extensive-dependence}. 
Since of $A_{ns}$ and $A_{ws}$ are identified
explicitly from from heterogeneity in an intensive property, $\gamma_s$, all associated energy will be embedded into the fluctuation term for that intensive property. 
Note that Eq. \ref{eq:fluctuation-wetting-v} is a simplification of the more general
case, where $\gamma_s$ may vary in complex ways along the grain surface, or change with time due transient chemical effects. 
We now consider particular geometric considerations for water-wet systems. 

\subsection{Geometric constraints}

As fluid is injected into a porous material, the configuration of fluids within the microstructure
changes, with interface rearrangement occurring alongside displacement. Changes to the fluid volume fraction will 
typically also lead to changes in interfacial area. However, since interfacial area can change without
any change to the volume fraction, the chain rule cannot be applied to simplify Eq. \ref{eq:entropy-inequality},
\[
\frac{\partial \overline{A}_n}{\partial t} \neq \frac{\partial \overline{A}_n}{\partial \overline{\phi}_w} 
\frac{\partial \overline{\phi}_w}{\partial t} \;.
\]
The rate of change in surface area is neither independent from the rate of change in the volume fraction,
nor entirely dependent on it. To obtain a flux-force form of the entropy inequality, the dependent component
must be distinguished from the independent component. 
It has been previously showed that a global geometric relationship can be constructed based on
a non-dimensional relationship developed based on invariant measures of integral geometry \cite{McClure_etal_2020}. 
Here we apply this relationship to the time-and-space averaged geometric measures
\begin{equation}
    f(\overline{\phi}_n, \overline{W}_n, \overline{X}_n) = 0 \;.
    \label{eq:geometric-state}
\end{equation}
The non-dimensional measures are defined as
\begin{equation}
\overline{W}_{n} = \frac{\overline{A}_n^2}{\overline{H}_n V}\;, \quad 
X_{n} = \frac{\overline{\chi}_n \overline{A}_n}{ \overline{H}_n^2}\;.
\label{eq:measures}
\end{equation}
where $\overline{A}_n$ is the time average for the surface area of the non-wetting fluid boundary,
$\overline{H}_n$ is the time average of integral mean curvature, and $\overline{\chi}_n$ is the
time average for the Euler characteristic. Even though the instantaneously measured
${\chi}_n$ is a discontinuous function of time, $\overline{\chi}_n$ will be continuous
due to its construction as a time average. While the Gibbs dividing surface can be considered
as introducing non-smooth aspects into the system description based on the discrete representation
of fluid regions according to set theory, the time-and-space average removes any non-smooth
behaviors by incorporating the past and future state of the system within the instantaneous
representation. To link the time derivatives for the surface area
and volume fraction, we note that a differential form is implied by Eq. \ref{eq:geometric-state}
\begin{equation}
 \frac{\partial \overline{\phi}_n}{\partial t} =
  \frac{\partial \overline{\phi}_n}{\partial \overline{W}_n}\frac{\partial \overline{W}_n}{\partial t}
  + \frac{\partial \overline{\phi}_n}{\partial \overline{X}_n}\frac{\partial \overline{X}_n}{\partial t} \;.
    \label{eq:geometric-state-diff}
\end{equation}
Since $\overline{W}_n$ and $\overline{X}_n$ include $\overline{A}_n$, this can be arranged to 
remove the rate of change in surface area from Eq. \ref{eq:entropy-inequality}.
The link between thermodynamics and the geometric evolution can be understood by
expanding the derivatives using the definitions in Eq. \ref{eq:measures}
\begin{eqnarray}
\frac{\partial \overline{W}_n}{\partial t} &=& \frac{\partial \overline{W}_n}{\partial \overline{A}_n}\frac{\partial \overline{A}_n}{\partial t}
    + \frac{\partial \overline{W}_n}{\partial \overline{H}_n}\frac{\partial \overline{H}_n}{\partial t} \nonumber \\
&=&  \frac{2 \overline{A}_n}{\overline{H}_n V} \frac{\partial \overline{A}_n}{\partial t} 
     - \frac{\overline{A}_n^2}{\overline{H}_n^2 V} \frac{\partial \overline{H}_n}{\partial t} \;.
    \label{eq:dWdt}
\end{eqnarray}
Topological contributions arise based on the time derivative of $\overline{X}_n$
\begin{eqnarray}
\frac{\partial \overline{X}_n}{\partial t} &=&\frac{\partial \overline{X}_n}{\partial \overline{A}_n}\frac{\partial \overline{A}_n}{\partial t}
    + \frac{\partial \overline{X}_n}{\partial \overline{H}_n}\frac{\partial \overline{H}_n}{\partial t}
     + \frac{\partial \overline{X}_n}{\partial \overline{\chi}_n}\frac{\partial \overline{\chi}_n}{\partial t}\nonumber \\
&=&   \frac{\overline{\chi}_n}{\overline{H}_n^2} \frac{\partial \overline{A}_n}{\partial t}
    - 2\frac{\overline{\chi}_n \overline{A}_n}{ \overline{H}_n^3} \frac{\partial \overline{H}_n}{\partial t}
    + \frac{ \overline{A}_n}{ \overline{H}_n^2}\frac{\partial \overline{\chi}_n}{\partial t} \;.
    \label{eq:dXdt}
\end{eqnarray}
Inserting the previous two expressions into Eq. \ref{eq:geometric-state-diff} and rearranging
terms leads to an expression that links the time derivatives 
\begin{eqnarray}
&& \frac{\partial \overline{\phi}_n}{\partial t} =
     + g_1 \frac{\partial \overline{A}_n}{\partial t} 
     -  g_2 \frac{\partial \overline{H}_n}{\partial t} 
     + g_3 \frac{\partial \overline{\chi}_n}{\partial t}
     \;,
      \label{eq:geometric-state-diff-expand}
\end{eqnarray}
where three geometric functions have been determined to depend on the geometric state of the system
\begin{eqnarray}
g_1 &=&  
%\Big(\frac{\partial f}{\partial \overline{\phi}_n} \Big)^{-1} 
\frac{2\overline{A}_n}{\overline{H}_n V} \frac{\partial \overline{\phi}_n}{\partial \overline{W}_n}
+  \frac{\overline{\chi}_n}{\overline{H}_n^2}\frac{\partial \overline{\phi}_n}{\partial \overline{X}_n} \;, \\
g_2 &=& 
     \frac{\overline{A}_n^2}{\overline{H}_n^2 V} \frac{\partial \overline{\phi}_n}{\partial \overline{W}_n} 
     + 2  \frac{\overline{\chi}_n \overline{A}_n}{ \overline{H}_n^3}\frac{\partial \overline{\phi}_n}{\partial \overline{X}_n}\;, \\
g_3 &=& \frac{ \overline{A}_n}{ \overline{H}_n^2} \frac{\partial \overline{\phi}_n}{\partial \overline{X}_n} \;.
\end{eqnarray}
The average curvatures for the surface can be obtained by dividing the integral mean curvature
and Euler characteristic by the total surface area,
\begin{equation}
 \overline{J}_n \equiv \frac{\overline{H}_n}{ \overline{A}_n}\;, \quad
 \overline{K}_n \equiv \frac{\overline{\chi}_n}{ \overline{A}_n}\;.
\end{equation}
This means that 
\begin{eqnarray}
\frac{\partial\overline{H}_n }{\partial t} &=& 
    \overline{J}_n \frac{\partial\overline{A}_n }{\partial t} + \overline{A}_n\frac{\partial\overline{J}_n }{\partial t} \;,  \\
\frac{\partial\overline{\chi}_n }{\partial t} &=& 
    \overline{K}_n \frac{\partial\overline{A}_n }{\partial t} + \overline{A}_n\frac{\partial\overline{K}_n }{\partial t} \;,
\end{eqnarray}
where the average mean curvature $\overline{J}_n$ and average Gaussian curvature $\overline{K}_n$
are now cast as intensive properties of the surface. Inserting into Eq. \ref{eq:geometric-state-diff-expand}
and rearranging terms leads to
\begin{eqnarray}
\frac{\partial \overline{\phi}_n}{\partial t} =
      \frac{\partial \overline{A}_n}{\partial t} ( g_1 - g_2\overline{J}_n + g_3 \overline{K}_n) && 
     \nonumber \\
     +  \overline{A}_n \Big ( g_3 \frac{\partial\overline{K}_n }{\partial t} 
     -    g_2 \frac{\partial\overline{J}_n }{\partial t}  \Big)
   \;,
      \label{eq:geometric-state-diff-expand-ii}
\end{eqnarray}
expanding terms we find that 
\begin{eqnarray}
    g_2\overline{J}_n 
%    &=&   
%    \overline{J}_n 
%    \Big(  \frac{\overline{A}_n^2}{\overline{H}_n^2 V} \frac{\partial \overline{\phi}_n}{\partial \overline{W}_n} 
%     + 2  \frac{\overline{\chi}_n \overline{A}_n}{ \overline{H}_n^3}\frac{\partial \overline{\phi}_n}{\partial \overline{X}_n}\Big)
%\nonumber \\
    &=&   \frac{1}{\overline{J}_n V}  \frac{\partial \overline{\phi}_n}{\partial \overline{W}_n} 
    + 2  \frac{\overline{\chi}_n}{ \overline{H}_n^2}\frac{\partial \overline{\phi}_n}{\partial \overline{X}_n}
\end{eqnarray}
and 
\begin{eqnarray}
    g_3 \overline{K}_n 
%    &=&   \overline{K}_n 
%    \Big(\frac{ \overline{A}_n}{ \overline{H}_n^2} \frac{\partial \overline{\phi}_n}{\partial \overline{X}_n} \Big)
%\nonumber \\
    &=& \frac{ \overline{\chi}_n}{ \overline{H}_n^2} \frac{\partial \overline{\phi}_n}{\partial \overline{X}_n} \;,
\end{eqnarray}
which means that 
\begin{eqnarray}
g_1 - g_2\overline{J}_n + g_3 \overline{K}_n &=& 
\frac{1}{\overline{J}_n V} \frac{\partial \overline{\phi}_n}{\partial \overline{W}_n} \;.
\end{eqnarray}
Solving to eliminate the time derivative for the surface area as it appears in the
thermodynamic expressions, 
\begin{eqnarray}
\frac{\partial \overline{A}_{n}}{\partial t} 
&=&  V \overline{J}_n \frac{\partial \overline{W}_n}{\partial \overline{\phi}_n}
\frac{\partial \overline{\phi}_n}{\partial t} 
+ \frac{\overline{A}_{n}}{ \overline{J}_n} \frac{\partial\overline{J}_n }{\partial t} \nonumber \\
&& 
- \frac{V}{ \overline{J}_n} \frac{\partial \overline{W}_n}{\partial \overline{X}_n}
\Bigg( 
     2 \frac{\overline{K}_n}{ \overline{J}_n} \frac{\partial\overline{J}_n }{\partial t}  
         - \frac{\partial\overline{K}_n }{\partial t}
          \Bigg) \;.
   \label{eq:geometric-state-diff-expand-iii}
\end{eqnarray}

\subsection{Flux-force form of entropy inequality}

Classical non-equilibrium theory relies on a flux-force form of the entropy
inequality to derive phenomenological equations \cite{deGroot_Mazur_84}.
Using the fact that the molecular system is time-reversible,
detailed balance is assumed \cite{Onsager_1931a}.
However, in systems with cooperative effects, such as interface rearrangements, detailed balance
will not hold within sub-regions of the system occupied by individual fluids. The fluctuation
terms are of central importance to this effect, since they capture the net contribution 
due to cooperative effects over $\Lambda$ and $\Omega$. As long as the fluctuations do 
not undermine the independence of terms in the flux-force entropy inequality, 
phenomenological equations may be derived. Using the results of the previous sections, we now obtain such a result. We consider situations where changes to the film thickness are caused by changes to the fluid volume fraction, 
\begin{equation}
    \frac{\partial \overline{h}_s}{\partial t} 
    = \frac{\partial \overline{h}_s}{\partial \overline{\phi}_w}
       \frac{\partial \overline{\phi}_w}{\partial t} \;.
       \label{eq:area-film-volume}
\end{equation}
Making use of Eqs. \ref{eq:geometric-state-diff-expand-iii}
and \ref{eq:area-film-volume} the entropy inequality in Eq. \ref{eq:entropy-inequality} can be put into
a flux-force form
\begin{eqnarray}
       \frac {\partial \overline{S}}{\partial t}
        &=& 
    \frac{V}{\overline{T} } \Bigg \{ \frac L V \Big[
  \underbrace{ \nabla \overline{\Phi}_w \cdot \overline{\mathbf{Q}}_w
    + \nabla \overline{\Phi}_n \cdot \overline{\mathbf{Q}}_n }_{\mbox{Work of Darcy flow}} \Big]
\nonumber \\
&& 
+\frac{\partial \overline{\phi}_w }{\partial t} \Bigg [
\underbrace{    \overline{p}_w - \overline{p}_n 
+ \overline{\gamma}_{wn} \overline{J}_n \frac{\partial \overline{W}_n}{\partial \overline{\phi}_n}
+ \frac{\overline{A}_s \overline{\Pi}_s}{V}\frac{\partial \overline{h}_{s}  }{\partial \overline{\phi}_w }
%-\overline{\gamma}_{wn} \frac{\partial \overline{a}_{wn}  }{\partial \overline{\phi}_w } 
}_{\mbox{capillary pressure dynamics}} \Bigg]
\nonumber \\
&&
+ \underbrace{
\Big<\frac {A_s }{V} \Big( \frac{\partial{\gamma}_s^\prime}{\partial t}
- h_s \frac{\partial{\Pi}_s^\prime}{\partial t}\Big)
-\phi_w \frac{\partial{p}_w^\prime}{\partial t} 
- \phi_n \frac{\partial{p}_n^\prime}{\partial t}\Big>
%\overline{\phi}_w \overline{p}_w^{\prime(0)} 
%- \overline{\phi}_n \overline{p}_n^{\prime(0)}
%+ \frac{\overline{A}_s \overline{h}_{s}}{V} \overline{\Pi}_{s}^{\prime(0)} 
 }_{\mbox{fluctuations and wetting dynamics}}
  \nonumber \\
&&
 - \underbrace{\frac{\overline{\gamma}_{wn} \overline{A}_{n} }{ \overline{J}_n V} \frac{\partial\overline{J}_n }{\partial t} 
+ \frac{\overline{\gamma}_{wn}}{ \overline{J}_n} \frac{\partial \overline{W}_n}{\partial \overline{X}_n}
\Bigg( 
     2 \frac{\overline{K}_n}{ \overline{J}_n}  \frac{\partial\overline{J}_n }{\partial t} 
          -  \frac{\partial\overline{K}_n }{\partial t} \Bigg)
          }_{\mbox{surface energy changes due to curvature}}
\Bigg\} \;.
  \nonumber \\
% Note -- sign for capillary pressure term switches because w phase volume fraction is used here instd of n in Eq. 34
% Note -- n pressure fluctuation terms multiply rate of change in n volume fraction also
\label{eq:flux-force-i}
\end{eqnarray}
Provided that $\Omega$ and $\Lambda$ are sufficiently large, the averaged quantities will be smooth functions of time,
since these are time averaged quantities. 
The fast dynamics will be entirely embedded within the associated fluctuation terms. 
Eq. \ref{eq:flux-force-i} is not yet in a flux-force form due to the contribution from fluctuations and the geometric evolution terms. Physically, this aligns with transient redistribution of energy within the system. 
For example, there may be reversible energy transfer from the pressure fluctuation terms to the surface energy. Note that a flux-force form can be obtained if
\begin{eqnarray}
\Big< \frac {A_s }{V} \Big( \frac{\partial{\gamma}_s^\prime}{\partial t}
- h_s \frac{\partial{\Pi}_s^\prime}{\partial t}\Big)
-\phi_w \frac{\partial{p}_w^\prime}{\partial t} 
- \phi_n \frac{\partial{p}_n^\prime}{\partial t}\Big>
%\overline{\phi}_w \overline{p}_w^{\prime(0)} 
%- \overline{\phi}_n \overline{p}_n^{\prime(0)}
%+ \frac{\overline{A}_s \overline{h}_{s}}{V} \overline{\Pi}_{s}^{\prime(0)} 
 - \frac{\overline{\gamma}_{wn} \overline{A}_{n} }{ \overline{J}_n V} \frac{\partial\overline{J}_n }{\partial t} 
 \nonumber \\
+ \frac{\overline{\gamma}_{wn}}{ \overline{J}_n} \frac{\partial \overline{W}_n}{\partial \overline{X}_n}
\Bigg( 
     2 \frac{\overline{K}_n}{ \overline{J}_n}  \frac{\partial\overline{J}_n }{\partial t} 
          -  \frac{\partial\overline{K}_n }{\partial t} \Bigg) =0 \;. \quad \quad \quad
\label{eq:fluctuation-constraint-ii}
\end{eqnarray}
%It would be worthwhile to further examine the possibility that each of the first and second line are independently zero. This will be left for future work. 
This expression defines a condition for the representative elementary volume (REV) for two-fluid flow through porous media. If terms on the left-hand side of Eq. \ref{eq:fluctuation-constraint-ii} do not sum to zero, then they constitute
is a net contribution from fluctuations to the energy dynamics. If
this happens, terms
on the first two lines of Eq. \ref{eq:flux-force-i} will not be independent
due to couplings caused by the internal energy dynamics. 
Assessment of terms in Eq. \ref{eq:fluctuation-constraint-ii} is therefore fundamental to the validity of macroscopic equations for two fluid flow.

To evaluate Eq. \ref{eq:fluctuation-constraint-ii} in experiments, pressure fluctuations need to be evaluated for both wetting and non-wetting fluids. The geometric state variables are needed to calculate the time derivatives for curvature terms and the slope of the geometric state function. We note that the fluctuations do not necessarily need to be Gaussian to obtain a flux-force entropy inequality. As long as any asymmetry in the fluctuations is  balanced by changes to the surface energy based on the evolution of interface configuration then Eq. \ref{eq:fluctuation-constraint-ii} can be satisfied.
Fluctuations should be considered from two perspectives: 
(1) steady-state flow at constant fluid volume fraction;
and (2) unsteady flow where the the fluid volume fraction is changing. 
In a steady-state flow, the curvature must be independent of time, since any net change to the geometric configuration
of fluids is inconsistent with the system being at steady-state. Therefore, the associated conditions
are 
\begin{eqnarray}
\frac{\partial\overline{K}_n }{\partial t} = 0\;, \quad \frac{\partial\overline{J}_n }{\partial t} = 0\;,
\\
%\overline{\phi}_w \overline{p}_w^{\prime(0)} + \overline{\phi}_n \overline{p}_n^{\prime(0)}
%-\frac{\overline{A}_s \overline{h}_{s}}{V} \overline{\Pi}_{s}^{\prime(0)}
\Big< \frac {A_s }{V} \Big( \frac{\partial{\gamma}_s^\prime}{\partial t}
- h_s \frac{\partial{\Pi}_s^\prime}{\partial t}\Big)
-\phi_w \frac{\partial{p}_w^\prime}{\partial t} 
- \phi_n \frac{\partial{p}_n^\prime}{\partial t}\Big>&=& 0 \;.
\label{eq:steady-fluctuation}
\end{eqnarray}
There must be no net rate of energy change from fluctuations at steady-state, since net changes to the system are inconsistent with the basic  notion
of steady state. For unsteady displacement, capillary fluctuations
may be balanced by changes to the surface energy.
This reflects the fact that the fluid configuration can change at constant fluid volume fraction based on 
changes to the boundary curvature. Since the boundary curvature can evolve independently from the volume,
a separate term must appear to account for these effects. 
 
Since the terms in Eq. \ref{eq:fluctuation-constraint-ii} do not multiply the change
in fluid volume fraction in Eq. \ref{eq:flux-force-i}, it is essential that their average value
be zero. Otherwise, the flux-force form of the entropy inequality cannot
be exploited to derive phenomenological equations. Any net contribution from the fluctuation terms
during steady-state flow invalidates any corresponding measurement of the the relative permeability.
On the other hand, if Eq. \ref{eq:fluctuation-constraint-ii} is satisfied then the effective permeability
coefficients can be considered to fully account for the dissipation,
including contributions from pore-scale events on the basis that the average rate of events is constant based on the averaging domain, $\Lambda$ and $\Omega$. Results obtained using pore-network models for steady-state fractional flow suggest that 
the associated fluctuations will be symmetric \cite{Winkler_etal_2019}.
In this case a flux-force form is  possible for Eq. \ref{eq:flux-force-i}, with the thermodynamic forces identified as
\begin{eqnarray}
\nabla \overline{\Phi}_w  \;, \quad
 \nabla \overline{\Phi}_n  \;, \quad
\frac{\partial \overline{\phi}_w }{\partial t} \;.
\end{eqnarray}
The resulting flux-force form of the entropy inequality can naturally be applied to derive
the standard constitutive relationship for relative permeability, given in Eq. \ref{eq:kr-sw}.
This conventional relationship assumes that 
the relationship between the flow rate $\overline{\mathbf{q}}_i$ and the 
applied forces is linear, which is known to break down as
the capillary number increases. This result is also based on the assumption that the forces and 
fluxes are independent, which has been called into question based on the results of
homogenization theory \cite{Whitaker_1986,Gray_Miller_14}. Previous efforts
to derive Eq. \ref{eq:kr-sw} from first principles have generally concluded that the 
flow rates $\overline{\mathbf{q}}_w$ and $\overline{\mathbf{q}}_n$ are not independent 
due to momentum exchanges between the two fluids. For example, if an external potential gradient
is applied only to one fluid, it will pull some amount of the other fluid along with it, 
which is not accounted for based on Eq. \ref{eq:kr-sw}. Cross-coupling forms have been proposed
to account for this effect \cite{Whitaker_1986,Flekkoy_1999}. 
However, since cross-coupling forms introduce additional 
phenomenological parameters, it is considerably more complicated to measure experimentally
and therefore difficult to apply in real systems. From the more practical standpoint, one can consider relative permeability as being a proxy for the actual dissipation rate, which 
depends non-linearly on the fluid geometry, the mobility ratio
capillary number, and other factors \cite{Lenormand_1988}. Since entropy is a scalar, 
we can always embed this rate within a scalar parameter. More recent efforts align with this more practical conceptual picture, and can be considered as characterizing the non-linear dependence of the actual dissipation rate on the applied forces \cite{Purswani2019}. In this context the fluctuation constraint in Eq. \ref{eq:steady-fluctuation} is a criterion that must be 
satisfied to accurately measure the steady-state dissipation, since this ensures that the energy
dynamics are stationary.

The rate of change for the fluid volume fraction has been previously used to derive
expressions for capillary pressure dynamics \cite{Hassanizadeh_Gray_93,Gray_Dye_etal_15}. 
We obtain an alternate form that includes the effect of both films and capillary fluctuations
\begin{equation}
     \frac{\partial \overline{\phi}_w }{\partial t}  = \frac {1}{\tau_w}
     \Bigg( 
 \overline{p}_w - \overline{p}_n 
+ \overline{\gamma}_{wn} \overline{J}_n \frac{\partial \overline{W}_n}{\partial \overline{\phi}_n}
+ \frac{\overline{A}_s \overline{\Pi}_s}{V}\frac{\partial \overline{h}_{s}  }{\partial \overline{\phi}_w }
%     \overline{p}_w - \overline{p}_n 
%+ \overline{\gamma}_{wn} \overline{J}_n \frac{\partial \overline{W}_n}{\partial \phi_w} 
%+\frac{\overline{\Pi}_s \overline{A}_{s}}{V} \frac{\partial \overline{h}_{s}  }{\partial \overline{\phi}_w }
%-F_c^* 
\Bigg) \;,
\label{eq:dynamic-pc}
\end{equation}
where $\tau_w$ is a phenomenological parameter.
%and $F_c^*$ accounts for the energy contribution
%that is due to capillary fluctuations 
%\begin{equation}
%F_c^* \equiv \sqrt{\frac{\rho D_p^3}{\gamma_{wn}}} 
% \Bigg( \phi_w \overline{p}_w^{\prime(*)} 
%- \phi_n \overline{p}_n^{\prime(*)} 
%- \frac{\overline{A}_{s} \overline{h}_{s}}{V} \overline{\Pi}_{s}^{\prime(*)} \Bigg) \;.
%\label{eq:fast-fluctuation}
%\end{equation}
Eq. \ref{eq:dynamic-pc} is an expression for the capillary pressure dynamics, which differs
from previous forms. First, the effect of film swelling is included,
which is important for the snap-off mechanism. The capillary term includes the contribution
over the entire fluid boundary, including the part of the boundary where the shape
is strongly influenced by the local solid microstructure. Due to the effect of films, 
the force-balance in these regions differs from the meniscus. In the averaged form, 
the equilibrium requirement is that all force contributions must cancel, including
the effect of the fluid pressures, meniscus curvature, and film contributions.
Averaging in both time and space ensures smoothness of the pressure signal with respect to time provided that the system size and time interval for averaging are sufficiently large. 
Second, the capillary fluctuation terms are accounted for explicitly based on
the REV constraint stated in Eq. \ref{eq:fluctuation-constraint-ii}. 
This REV constraint is the main assumption used to derive Eq. \ref{eq:dynamic-pc}.
It is possible that net contribution due to capillary fluctuations will vanish in REV-scale systems due to the cancellation of terms. From the
perspective of pore-scale events, $\Omega$ and $\Lambda$ must be large enough to capture the distribution of events within a particular material. This will depend on both the material and the flow rate. However, the fluctuation term is not guaranteed to vanish because the pressure fluctuation signal may be non-stationary during unsteady displacement. If this is the case,
then Eq. \ref{eq:flux-force-i} does not reduce to a flux-force form and
additional treatment would be required to derive macroscopic phenomenological equations.
It is therefore important to treat the fluctuation terms cautiously. In the following sections, we consider several practical examples to illustrate the behaviour for capillary fluctuations and their contribution to the energy dynamics in multiphase systems. 

\section{RESULTS}
\subsection*{Flow through porous media}

The flow of immiscible fluids in porous media are strongly influenced by geometric and topological effects \cite{Land_68,Lenhard_Parker_87,Herring_Harper_etal_13,McClure_Berrill_etal_16b,Liu_Herring_etal_17,McClure_Armstrong_etal_2018,Schlueter2016b,Hilfer_06,Armstrong_McClure_etal_16}. Capillary forces typically dominate such flows, which are directly impacted by
changes to fluid connectivity. 
We consider fluid displacement within a porous medium as a means to explore how geometric 
changes influence physical behavior. A periodic sphere packing
was used as the flow domain. Using a collective rearrangement algorithm, 
eight equally-sized spheres with diameter $D=0.52$ mm 
were packed into a fully-periodic domain $1 \times 1 \times 1$ mm in size.
The system was discretized to obtain a regular lattice with $256^3$ voxels. 
Simulations of fluid displacement were performed by injecting fluid into the system
so that the effect of topological changes could be considered in detail. 
Initially the system was saturated with water, and fluid displacement was then simulated based 
on a multi-relaxation time (MRT) implementation of the color lattice Boltzmann model \cite{mcclure2020lbpm}.

\begin{figure}[ht]
\centering
\includegraphics[width=1.0\linewidth]{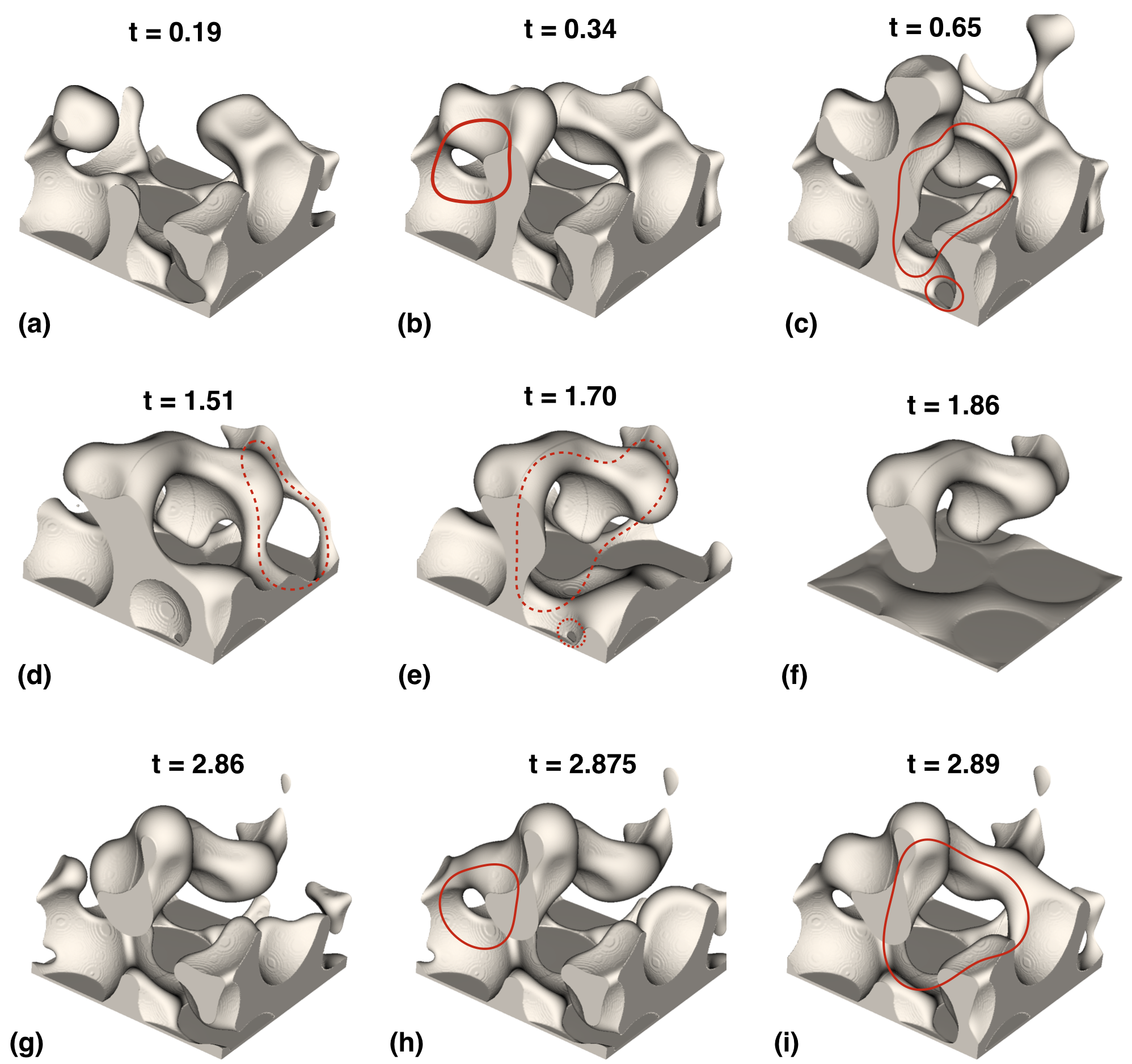}
\caption{Coalescence and snap-off events alter the topology
of fluids within porous media:
(a)-(c) pore-scale events such as Haines jumps lead to the formation of non-wetting fluid loops that 
are associated with coalescence events that occur during pore-filling;
(d)-(f) imibition destroys fluid loops due to snap-off, eventually trapping non-wetting fluid within the porespace; 
(h)-(i) trapped fluids reconnect during secondary drainage.}
\label{fig:porescale}
\end{figure}

To simulate primary drainage, non-wetting fluid
was injected into the system at a rate of $Q=50$ voxels per timestep,
corresponding to a capillary number of $4.7\times 10^{-4}$. A total of
500,000 timesteps were performed to allow fluid to invade the pores within the system, leading to the sequence of structures shown in Fig. 
\ref{fig:porescale}(a)--(c). The first pores are invaded at $t=0.19$ pore volumes,
which is just after the system overcomes the entry pressure. 
The fluid does not yet form any loops, since 
a pore must be invaded from two different directions before a loop will
be created. Haines jumps and loop creation can be considered as
distinct pore-scale events that may coincide in certain situations.
At time $t=0.34$ pore volumes the first loop is formed. As primary drainage progresses additional loops form, causing the Euler characteristic tends to decrease as non-wetting fluid is injected.

At the end of primary drainage the flow direction was reversed to induce imbibition, with
water injected at a rate of $Q = 20$ voxels per timestep. Sequences from the displacement
are shown in Fig. \ref{fig:porescale}(d)--(f). As water imbibes into the
smaller pores in the system, there is a well-known tendency to snap off non-wetting fluid by destroying connections to larger adjacent pores where imbibition has not yet occurred. This is known as the Roof snap-off mechanism \cite{Roof_1970}. Snap-off breaks loops
within the system, causing the Euler characteristic to increase. The order that loops
snap off during imibition is not the reverse of the order that the loops form during drainage. This is because the drainage process is dominated by the size of the throats whereas imibition is governed by the size of the pores. Eventually, the sequence of snap-off events can cause non-wetting fluid to become fully 
disconnected from the main region, leading to trapping as seen at time $t=1.86$ pore volumes.

\begin{figure}[ht]
\centering
\includegraphics[width=1.0\linewidth]{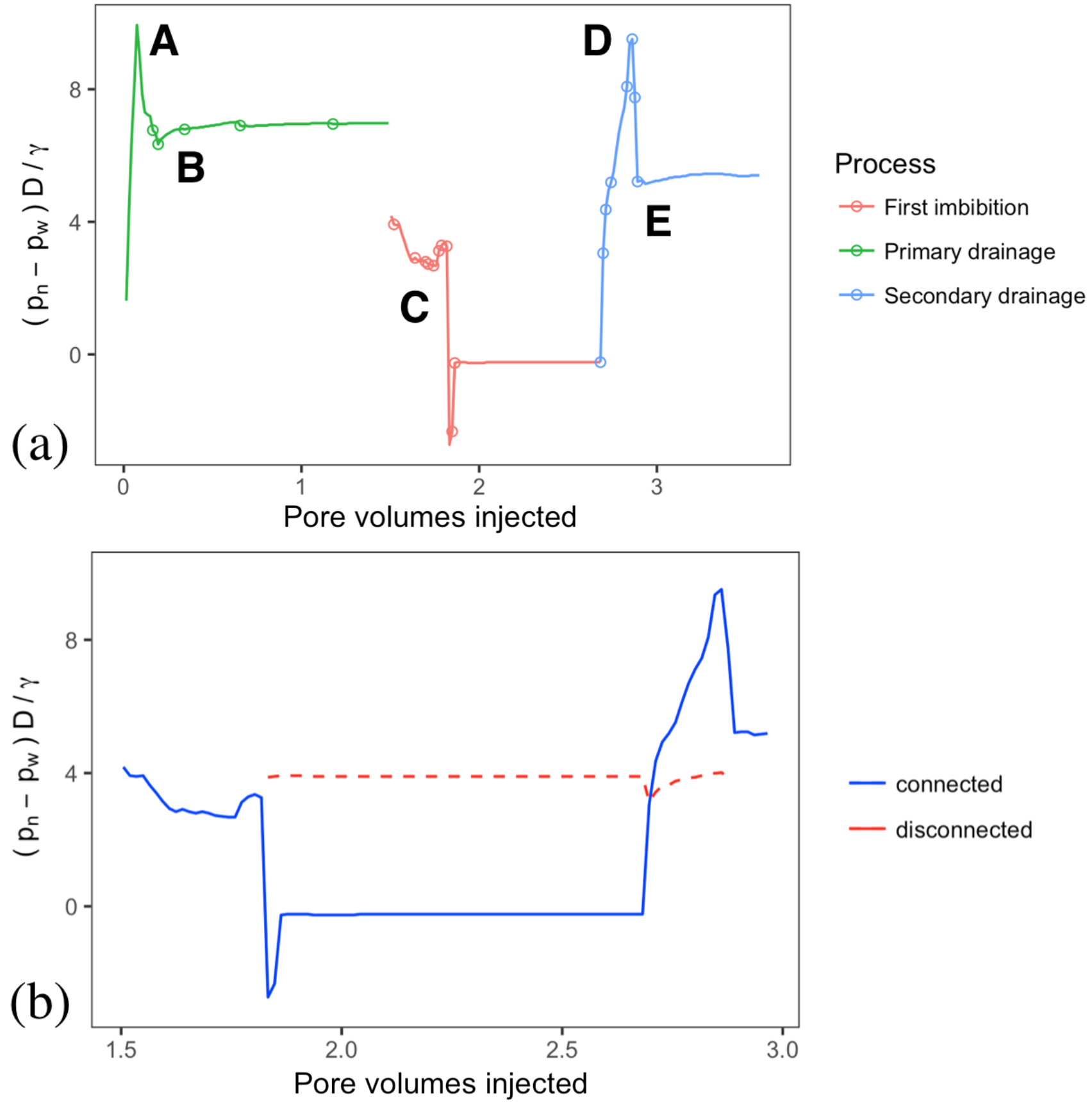}
\caption{Response of the capillary pressure to connectivity events:
(a) the difference between the pressure of the connected water and non-wetting
fluid show that non-smooth disruptions to the capillary pressure landscape align 
with connectivity events identified by changes in Euler characteristic (circles);
(b) snap-off causes large disruptions in capillary pressure, with non-wetting
fluid trapped at higher capillary pressure than the bulk fluid.}
\label{fig:pc}
\end{figure}

The presence of trapped fluid distinguishes secondary drainage from 
primary drainage. The sequence of fluid configurations shown in Fig. \ref{fig:porescale}(g)--(i) depicts 
the sequence of coalescence events that reconnect trapped fluid. 
Due to the presence of trapped fluid, the order that loops are formed
within the porespace does not match the order observed during primary 
drainage. Primary and secondary drainage each show a rapid spike in the pressure 
difference due to the pore entry pressure. The entry sequence along primary drainage
matches the configurations from Fig. \ref{fig:porescale}(a)--(c) and corresponds to events labeled
as A and B in Fig. \ref{fig:pc}(a).  The high pore entry pressure is consistent with expected behavior that the
largest energy barriers in the capillary dominated system are due to to
pore entry \cite{Berg_Ott_etal_13}. The pore entry sequence along
secondary drainage is shown in Fig. \ref{fig:porescale}(g)--(i) 
and labeled as D and E in Fig. \ref{fig:pc}(a).
On secondary drainage, the entry pressure is slightly reduced
due to the presence of trapped fluid. This shows that
the first coalescence event depicted in Fig. \ref{fig:porescale}(g)--(i) corresponds
exactly to the maximum fluid pressure difference observed during pore entry. 
This is distinct from the initial pore entry observed during primary drainage; 
upon secondary drainage a fluid coalescence event occurs instead of a Haines jump. 
The resulting pressure fluctuations differ as a consequence. 

Like coalescence, snap-off events are associated with clear fluctuations in
the bulk fluid pressures. The snap-off sequence shown in Fig. \ref{fig:porescale}(d)--(f) is
labeled as C in Fig. %\ref{fig:measures-porescale}e
\ref{fig:pc}(a). As loops of fluid are destroyed, corresponding
fluctuations in the fluid pressure difference are observed. The largest disruption is associated with the snap-off event.
The capillary pressure for the connected and disconnected regions of non-wetting fluid
is shown in Fig. %\ref{fig:measures-porescale}f
\ref{fig:pc}(b). Consistent with the observation of other authors,
fluid is trapped at a higher capillary pressure than what is measured in the connected
fluid phases at the instant of snap-off \cite{Roof_1970}.
As soon as snap-off occurs, the time derivative of the pressure difference between
the two connected fluid phases immediately jumps from strongly negative to strongly
positive. This occurs as each fluid region relaxes toward its preferred equilibrium 
state, which is possible because the two fluid regions are no 
longer mixing after the snap-off event. The final pressure for the trapped region is determined based on geometry, with the fluid assuming a minimum energy configuration within the pore
structure where it is trapped. 

\begin{figure}[ht]
\centering
\includegraphics[width=1.0\linewidth]{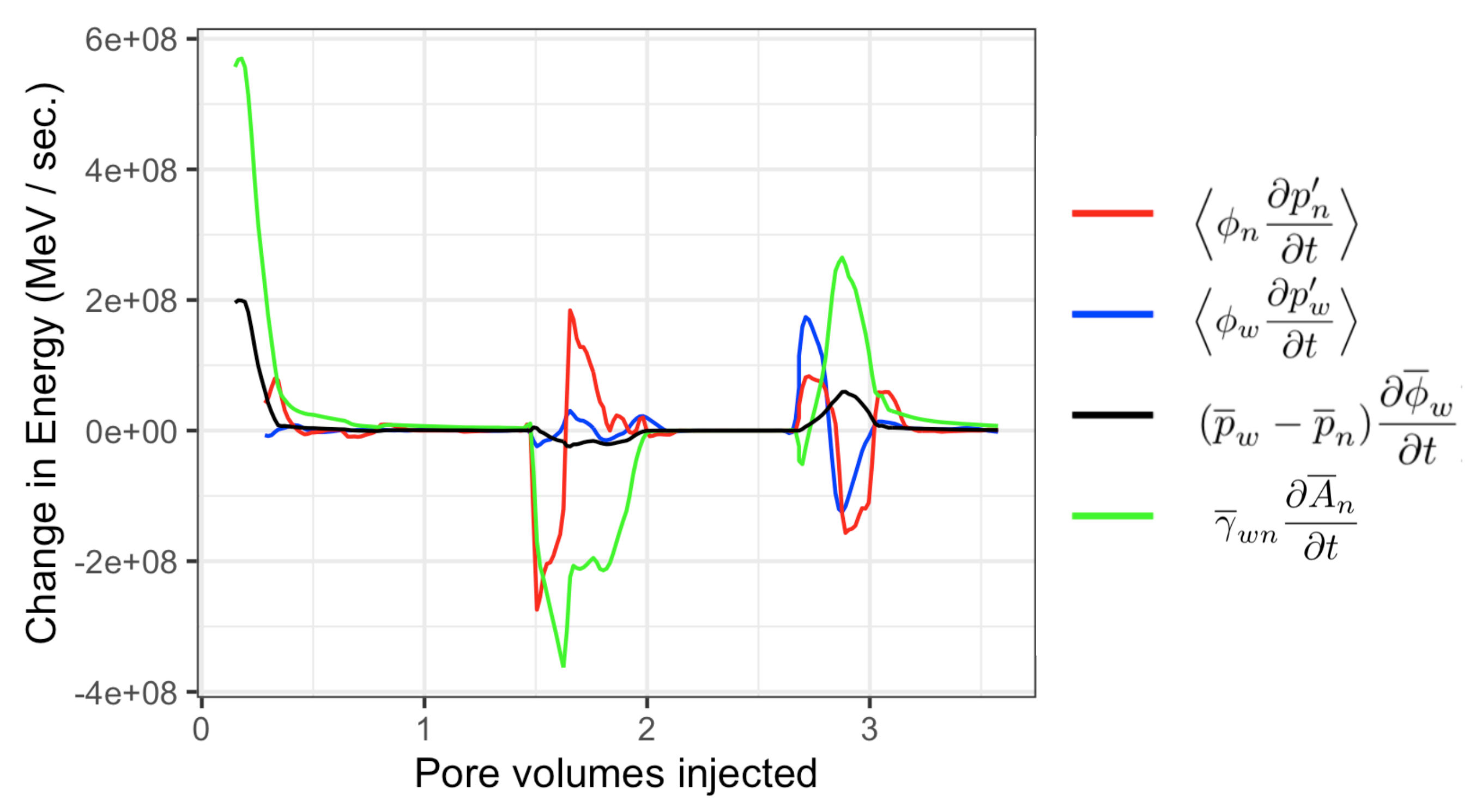}
\caption{Contribution of energy terms based on typical pore-scale events for 
drainage and imbibition in porous media. }
\label{fig:porescale-fluctuation}
\end{figure}

The energy associated with pressure fluctuations that arise due to pore-scale events 
is shown in Fig. \ref{fig:porescale-fluctuation}. At the pore level, the energy due
to fluctuations is comparable to the rate of pressure volume work and the rate of
change in surface energy. Defining the exchanges between these terms in a non-equilibrium
setting is non-trivial, since it is determined in part by the geometry of the problem. 
It is clear from these results that the fluctuation terms contribute a 
significant amount to the overall energy dynamics at the scale of an individual pore.
We now extend this analysis to larger scale systems where large numbers of pore-scale
events occur, allowing their frequency and distribution to be analyzed in detail. 

\subsection*{Experimental Results}

\begin{figure}[ht]
\centering
\includegraphics[width=1.0\linewidth]{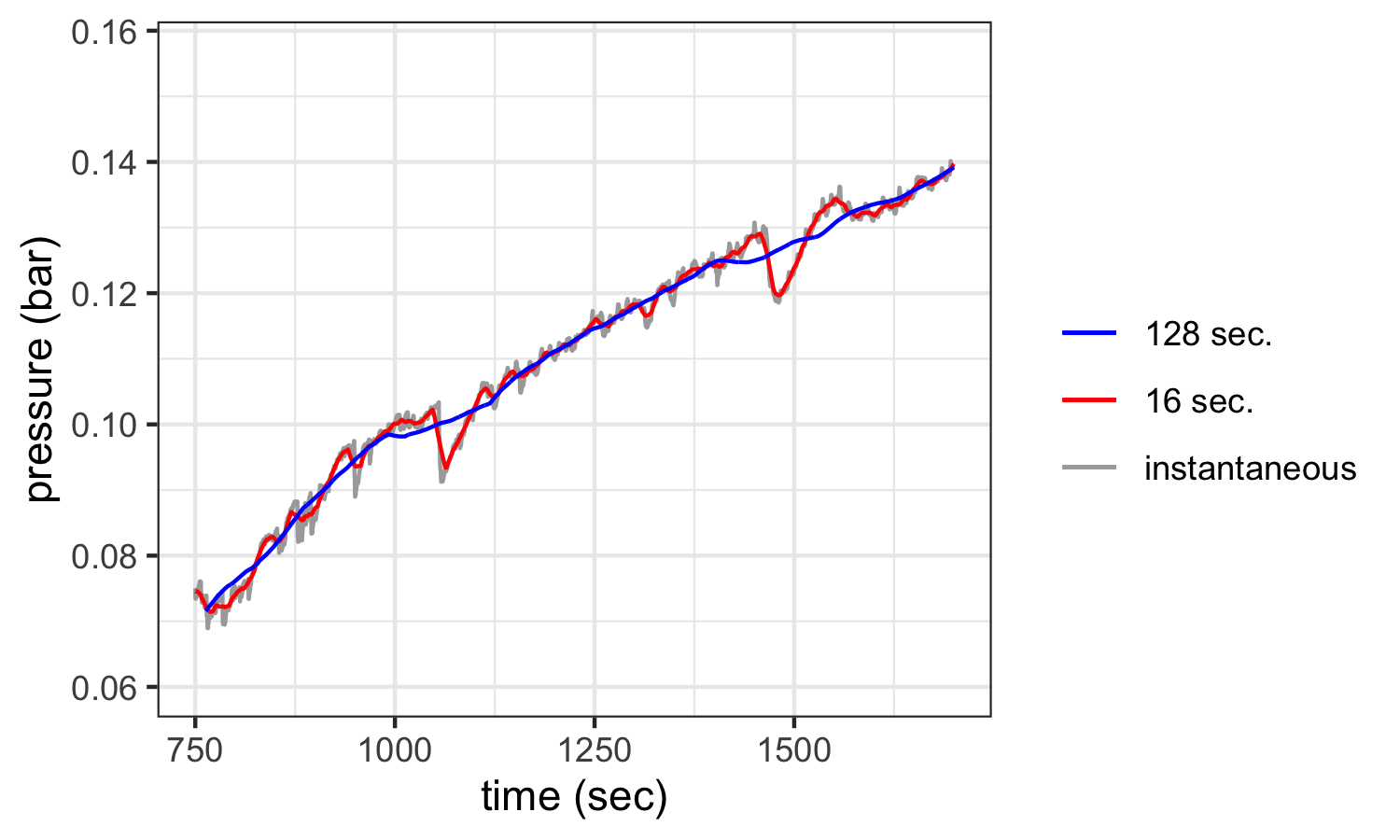}
\caption{Experimental pressure signal from primary drainage in a sandstone with time-averaged pressures obtained based on a time interval of 16 s. and 128 s. Time-averaging smooths
the fluctuations in the pressure signal; the associated contributions to the energy are linked
with the fluctuation terms.}
\label{fig:experimental-pressure}
\end{figure}
Experimental data was collected for a Berea sandstone.
The sample porosity was $0.199$ and the absolute permeability was $0.7$ $\mu$m$^2$.
The sample diameter was $4$ mm and the length $10$ mm.
Oil was injected into the sample at a rate of $0.35$ $\mu$L/minute. 
Pressure transducer measurements were collected for the oil at an interval
of $\Delta t = 0.32$ sec. Several assumptions were necessary to support the analysis of the experimental data. 
Since pressure measurements were available for the oil phase only,
the water was assumed to be in equilibrium with the external pressure 
(1 atm) at the outlet of the sample. Since the transducer measurements for the oil pressure are
available only at the inlet boundary, this was assumed to adequately represent the spatially averaged
pressure within the sample. While it is nearly certain that various intermittent pressure gradients 
and spatial fluctuations occurred within the sample, the response of the pressure waves (based on the speed of sound) is fast compared to the timescale for flow (based on the rate of change in saturation). Time derivatives were evaluated by fitting a spline function to the experimental time series data and evaluating the first derivative. Time averages were then computed based on a simple moving average (SMA). Average pressure values are shown in Fig. \ref{fig:experimental-pressure}. For $\Lambda=128$ seconds, the pressure signal behaves as a continuous {\em ison}, incorporating the reversible components of the original signal. Given a sufficiently long
time averaging interval, the effect of {\em rheons} is thereby embedded entirely within the fluctuation terms. 

\begin{figure}[ht]
\centering
\includegraphics[width=1.0\linewidth]{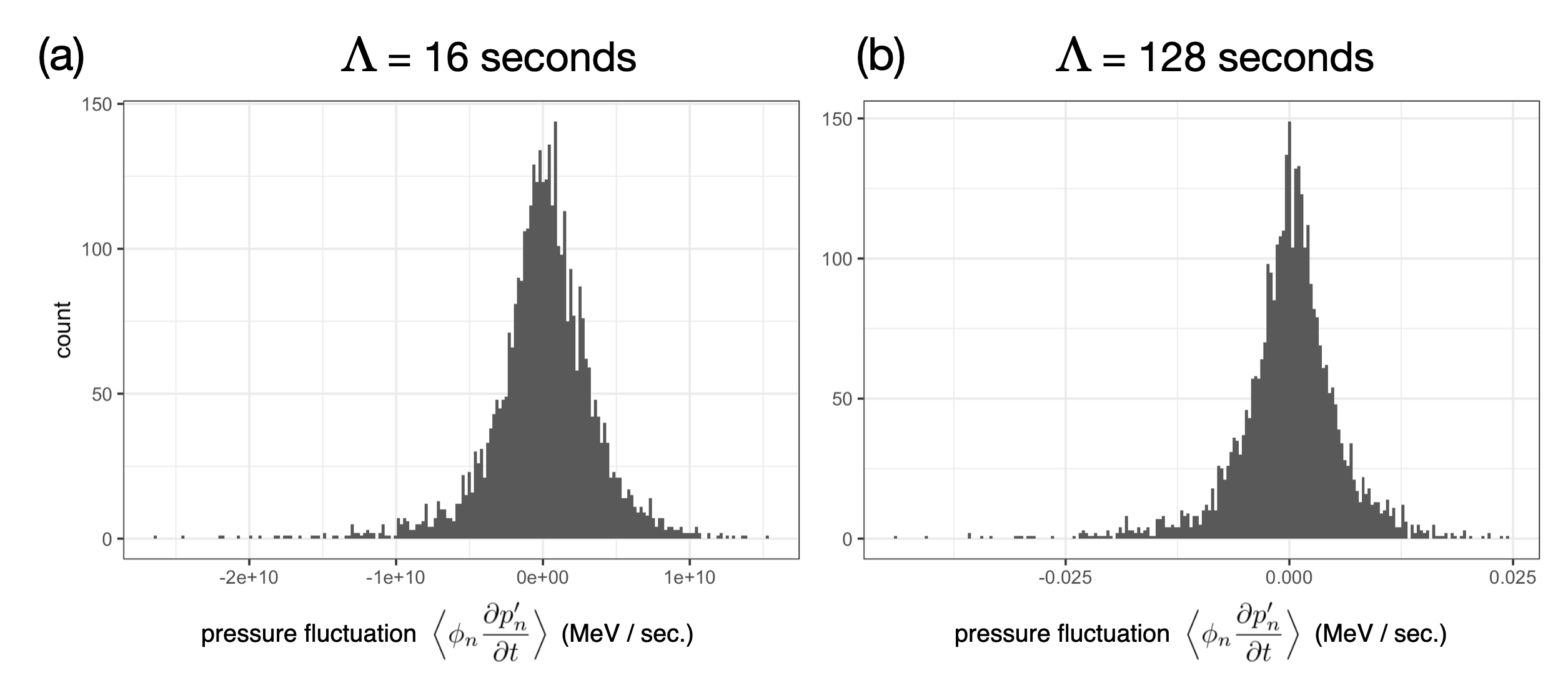}
\caption{Observed distribution for pressure fluctuation determined using  experimental data with the temporal averaging interval (a) $\Lambda=16$ sec. (b) $\Lambda=128$ sec. }
\label{fig:experimental-fluctuation-histogram}
\end{figure}

\begin{figure}[ht]
\centering
\includegraphics[width=1.0\linewidth]{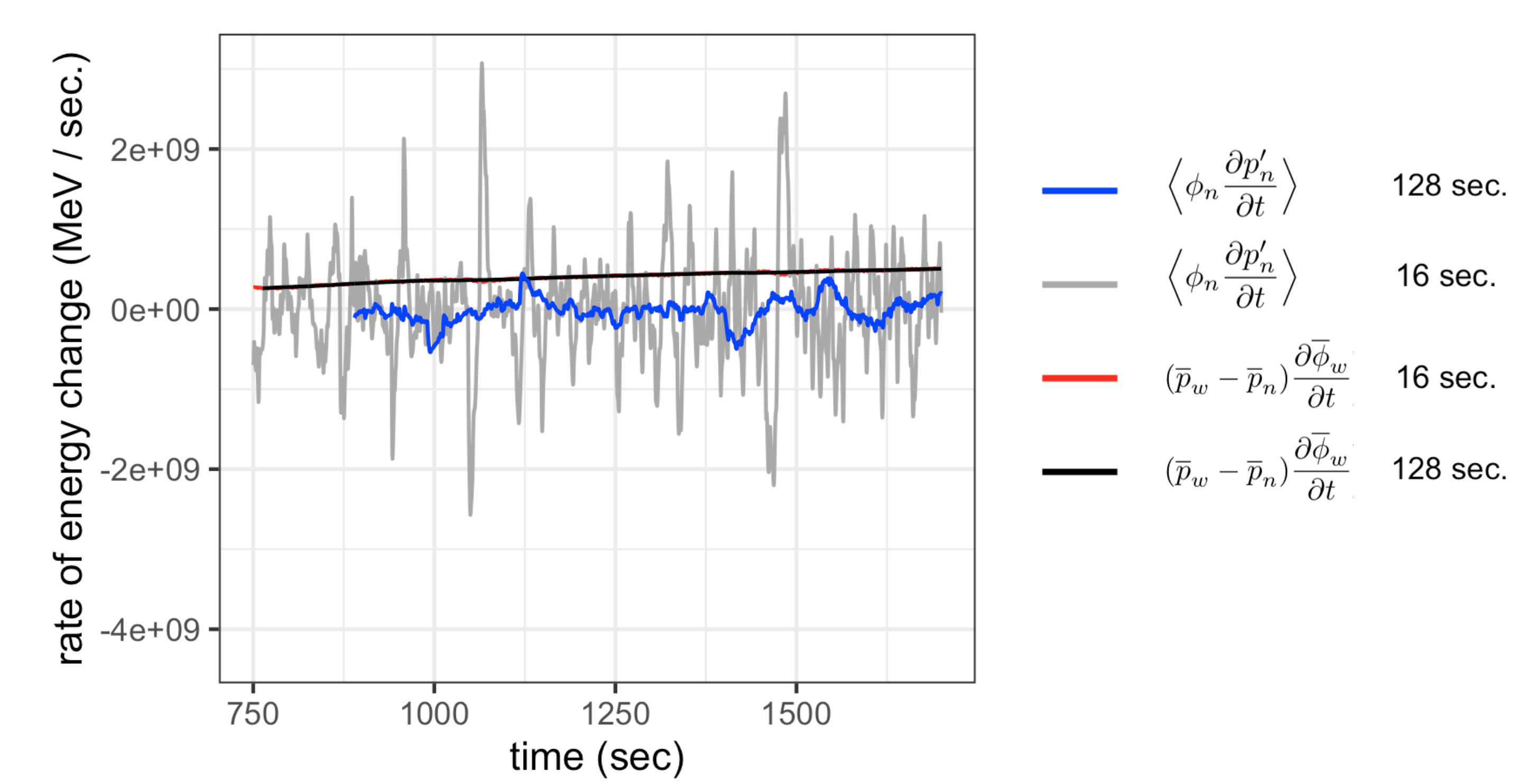}
\caption{The amplitude of pressure fluctuations is reduced based on the interval for 
time averaging. As the time interval for averaging is increased, fluctuations become
less significant as compared to the rate of pressure volume work.}
\label{fig:experimental-fluctuation}
\end{figure}

Fig. \ref{fig:experimental-pressure} shows the experimental pressure signal
and time-averaged curves computed based two time intervals, $\Lambda=16$ seconds and $\Lambda=128$ seconds. 
The associated fluctuation distributions are shown in Fig. \ref{fig:experimental-fluctuation-histogram}.
Pressure fluctuations continue to be evident due to the larger events with
$\Lambda=16$ sec. These fluctuations are mostly removed for $\Lambda=128$ sec, although the 
effect of larger events remains visually evident. Time average effectively removes the noise
due to the fast pressure fluctuations from the pressure signal. The distribution for the pressure fluctuation is shown in Fig. \ref{fig:experimental-fluctuation-histogram}. Since the pressure fluctuation is defined based on the difference between the instantaneous and average pressure values, the amplitude is sensitive to the averaging interval. Fig. \ref{fig:experimental-fluctuation-histogram}(a) shows
the distribution for $\Lambda = 16$ seconds. The distribution for  $\Lambda = 128$ seconds is shown in Fig. \ref{fig:experimental-fluctuation-histogram}(b).
According to the central limit theorem, the distribution should be well-approximated by the Gaussian distribution provided that $\Lambda$ is large enough to include a sufficiently large number of pore-scale events. Standard assumptions also require that the average pressure $\overline{p}_i$ be stationary; if the 
average pressure pressure value is drifting in time, then skew may be introduced in
the pressure fluctuation distribution. This is an effective condition for the REV of
the system. As the system grows larger, the rate of change to the fluid volume fraction
$\partial \overline{\phi}_w / \partial t$ will be slower, since it takes a longer time
for flow to fill the larger system. However, the timescale for pore-scale events will be the
same. The average pressure signal will tend to be more stationary for larger systems considered over longer time intervals. For a fixed system size, a larger $\Lambda$ will smooth the pressure signal and reduce the amplitude of the pressure fluctuations, as shown in Fig. \ref{fig:experimental-fluctuation}. However, a larger value of $\Lambda$ will mean the distribution is less stationary, since the average pressure can drift more over a a longer
time interval. For example, it is clear from  Fig. \ref{fig:experimental-pressure} that
the average pressure difference is changing at the timescale for $\Lambda=128$.

A reasonable criterion for the REV is that the magnitude
for the pressure fluctuations (based on the variance for the pressure fluctuation
distribution as depicted in Fig. \ref{fig:experimental-fluctuation-histogram}) must
be small as compared to the rate of pressure-volume work. If this condition is not met,
then {\em rheons} will exert an asymmetric influence on the pressure signal and
Eq. \ref{eq:fluctuation-constraint-ii} may not hold.
In this situation, the average energy landscape is ``rugged" in nature and the fluctuations
must be considered explicitly. In a sufficiently large system, considered over a sufficiently
large time interval, Eq. \ref{eq:fluctuation-constraint-ii} should hold and an averaged non-equilibrium description should be possible. Fig. \ref{fig:experimental-fluctuation} shows the amplitude of pressure fluctuations on the energy scale, which includes the rate of pressure-volume work. The instantaneous values for the pressure fluctuation can be very large, due to the fact that pore-scale events are very fast. The overall energy contribution is significant, and cannot be ignored unless REV requirements are met. An intriguing possibility is that various terms in Eq. \ref{eq:fluctuation-constraint-ii} may counter-balance asymmetry in the pressure
fluctuations. For example, pressure fluctuation energy can be redistributed to the 
surface energy, or to the other fluid. Part
of the energy from {\em rheons} can be reversibly transferred to other internal energy
modes.

\begin{figure}[ht]
\centering
\includegraphics[width=1.0\linewidth]{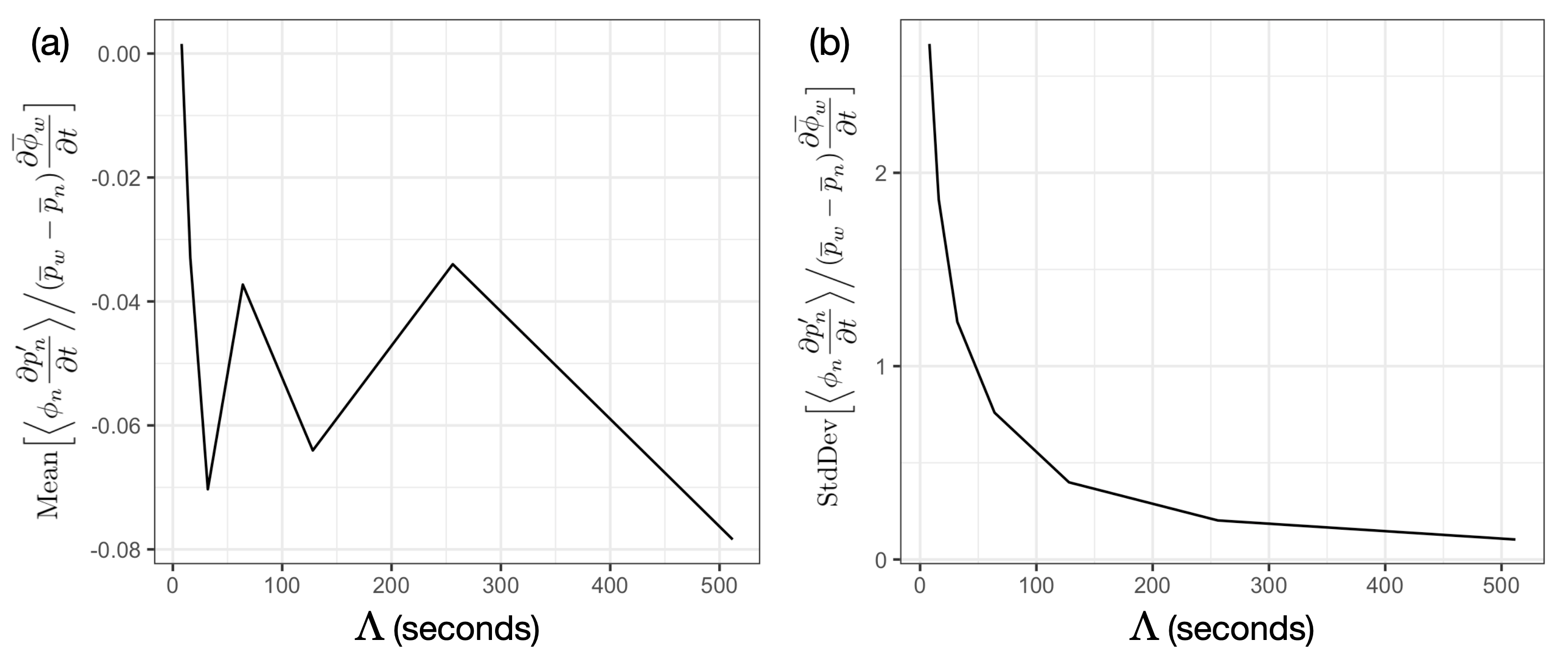}
\caption{Temporal REV assessment for experimental data by comparing the energy
due to fluctuations with the rate of pressure volume work: (a) due to the change in the 
capillary pressure and volume fraction over the averaging interval $\Lambda$, fluctuations 
have non-zero mean value. Fluctuations will tend to have mean value closer to zero in spatially larger systems or when $\Lambda$ is smaller;
(b) standard deviation for fluctuations decreases toward zero as the time
interval $\Lambda$ increases, such that many pore-scale events are factored into the averages. }
\label{fig:REV}
\end{figure}

An assessment of the temporal REV is shown in Fig. \ref{fig:REV}. Assessment of the REV
based on the pressure fluctuations must consider two essential components: 
(1) the mean for the pressure fluctuation distribution, since a non-zero mean implies
there is a net contribution to the internal energy;
(2) the variance associated with the amplitude of the pressure fluctuation, which is related
to the number of pore-scale events within the time interval $\Lambda$.
For the particular experimental data considered here, the energy associated with the 
mean pressure fluctuation is $\sim 5$ \% of the rate of pressure-volume work. This can 
likely only be further reduced by increasing the physical size of the sample. The 
standard deviation for the pressure fluctuation decreases as $\Lambda$ increases,
decreasing to $\sim 10$ percent of the rate of pressure-volume work for $\Lambda=512$
seconds. When considered over short time averaging intervals the fluctuations are several
times the rate of pressure volume work. Based on the metrics defined in Fig. \ref{fig:REV}
the REV accuracy for the experimental system considered is likely to be in the range of 
$\pm 5$--$10$ \% for the capillary pressure. 

It is noted that the requirement given
in Eq. \ref{eq:fluctuation-constraint-ii} can still be satisfied if the net contribution from 
the pressure fluctuation balances against the change in surface energy and the contribution
due to film swelling. However, this cannot be assessed for the data set considered here, since the data needed to evaluate the curvature and film contributions was not collected. Additional study of this relationship is important to establish the validity of macroscopic flow equations. 
Our results suggest that during unsteady flow, the pressure fluctuations within an individual
fluid will makes a net contribution to the energy dynamics that cannot be ignored. 
To meet an REV criterion these fluctuations must either cancel with other terms in
Eq. \ref{eq:fluctuation-constraint-ii}, or the net effect of fluctuations must vanish 
for sufficiently large systems. 

It is important to recognize that pore-scale events will always dissipate a significant
amount of energy during two-fluid flow in porous media. 
Averaging does not change this. What averaging does accomplish is to
treat fluctuations such that their energy dissipation is accounted for based on 
Eqs. \ref{eq:kr-sw} and \ref{eq:dynamic-pc}, i.e. in situations where
Eq. \ref{eq:fluctuation-constraint-ii} is satisfied.
The constitutive models for relative permeability and dynamic capillary pressure are
not valid if these constraints are not met. However, when the system size is large and long
time intervals are considered, the net contribution of the capillary fluctuation terms 
will trend towards zero. Their role on dissipation within the system will then be incorporated
into the phenomenological coefficients $k_{rn}$, $k_{rw}$ and $\tau_w$. 

\section*{Summary}

We derive non-equilibrium thermodynamics for immiscible fluid flow in porous media using the 
method of time-and-space averaging. Averaging in both time and space reveals
macroscopic thermodynamic fluctuation terms, which are athermal in nature and account
for internal energy transfers due to multi-scale effects such as capillary pressure dynamics. 
An entropy production equation is derived that includes the relevant fluctuation terms,
which is converted into flux-force form to support the derivation of phenomenological 
equations. The standard equations for effective permeability can be derived, which can be
generalized to cross-coupling forms in situations where significant momentum exchanges occur
between fluids. Key results of the theory are (1) a time-and-space REV constraint on fluctuations observed
during steady-state displacement, which is needed to derive the flux-force form of the entropy 
inequality; and (2) a dynamic capillary pressure expression that accounts for the effects of films and capillary fluctuations. Using simulation and experimental data we show that capillary fluctuations represent a significant part of the non-equilibrium energy dynamics for immiscible fluid flow. Based on experimental data, we assess the distribution for capillary fluctuations for time intervals ranging from $\Lambda=8$ seconds to $\Lambda=512$ seconds. The net contribution for fluctuations during unsteady displacement varies from $5$--$10$\% of external pressure-volume work. Our results indicate that capillary fluctuations are an essential consideration to determine REV for immiscible displacement. Further work is needed to determine whether net energy contributions due to capillary fluctuations and surface energy will cancel based on the REV constraint
derived in this paper.

%\subsection*{Acknowledgements}

\begin{acknowledgments}
An award of computer time was provided by the Department of Energy Summit Early Science program. This research also used resources of the Oak Ridge Leadership Computing Facility, which is a DOE Office of Science User Facility supported under Contract DE-AC05-00OR22725. $\mu$CT was performed on the TOMCAT beamline at the Swiss Light Source, Paul Scherrer Institut, Villigen, Switzerland. We are grateful to G. Mikuljan at Swiss Light Source, whose outstanding efforts have made these experiments possible'
\end{acknowledgments}

\section*{Data Availability Statement}

The data that support the findings of this study are available within the article and its supplementary material.

%\nocite{*}
%\bibliography{ergodic}% Produces the bibliography via BibTeX.

\begin{thebibliography}{100}

\bibitem{Armstrong_Berg_2013}
 {R.T. Armstrong and S. Berg},
 \newblock{``Interfacial velocities and capillary pressure gradients during Haines jumps"}.
 \newblock{Phys. Rev. E}, \textbf{88}, {043010}, (2013).
%  doi = {10.1103/PhysRevE.88.043010},
%  url = {https://link.aps.org/doi/10.1103/PhysRevE.88.043010}

\bibitem{Berg_Ott_etal_13}
S. Berg, H. Ott, S.A. Klapp, A. Schwing, R. Neiteler, N.
  Brussee, A. Makurat, L. Leu, F. Enzmann, J. Schwarz,
  M. Kersten, S. Irvine,  M. Stampanoni,
\newblock {``Real-time 3d imaging of Haines jumps in porous media flow,"}
\newblock { PNAS}, {\bf110}, 3755--3759, (2013).

\bibitem{Armstrong_Ott_etal_2014}
R.T. Armstrong, H. Ott, A. Georgiadis, M. R\"{u}cker, A. Schwing A, and S. Berg,
\newblock{``Subsecond pore-scale displacement processes and relaxation dynamics in multiphase flow."} 
\newblock{Water Resour Res} {\bf 50} 9162--9176 (2014).
%doi: 10.1002/2014WR015858.

\bibitem{Cueto-Felgueroso_Juanes_2015}
L. Cueto-Felgueroso, R. Juanes.
\newblock{``A discrete-domain description of multiphase flow in porous media: rugged energy landscapes and the origin of hysteresis"}
\newblock{Geophysical Research Letters} {\bf 43}, 1615--1622 (2015)

\bibitem{Morrow_1970}
N.R. Morrow,
\newblock{``Physics and thermodynamics of capillary action in porous media,"} 
\newblock{Ind. Eng. Chem} {\bf 63}, 32--56 (1970).

\bibitem{Primkulov_2019}
B. K. Primkulov, A.A. Pahlavan, X. Fu, B. Zhao, C.W. MacMinn and R. Juanes,.  
\newblock{``Signatures of fluid–fluid displacement in porous media: wettability, patterns and pressures,"}
\newblock{Journal of Fluid Mechanics} {\bf 875}, R4 (2019).
%doi: 10.1017/jfm.2019.554.


\bibitem{Winkler_etal_2019}
M. Winkler, M.a. Gjennestad, D. Bedeaux, S. Kjelstrup, R. Cabriolu, R. and A. Hansen, 
\newblock{``Onsager-Symmetry Obeyed in Athermal Mesoscopic Systems: Two-Phase Flow in Porous Media,"}     
\newblock{Frontiers in Physics}, {\bf 8},  {60} {(2020)}.      
%DOI={10.3389/fphy.2020.00060}


\bibitem{Gnesotto_2018}
{F. S. Gnesotto, F. Mura, J. Gladrow, and C.P. Broedersz},
\newblock{``{Broken detailed balance and non-equilibrium dynamics in living systems: a review},"}
\newblock{Reports on Progress in Physics},	{\bf 81} 066601 (2018).
%	doi = {10.1088/1361-6633/aab3ed},

\bibitem{Onsager_1931a}
L. Onsager, 
\newblock{``Reciprocal Relations in Irreversible Processes,"}
\newblock{Phys. Rev.} {\bf 37}, 405--426 (1931).

\bibitem{deGroot_Mazur_84}
S.R. de Groot and P. Mazur,
\newblock {{``Non-Equilibrium Thermodynamics,"}}
 Dover, Amsterdam, (1984).

\bibitem{Hassanizadeh_Gray_93}
S.M. Hassanizadeh and W.G. Gray,
\newblock{``Toward an improved description of the physics of two-phase flow,"}
\newblock { Adv. Water Res.}, {\bf16}, 53--67 (1993).

\bibitem{Kjelstrup_2019}
S. Kjelstrup, D. Bedeaux, A. Hansen, B. Hafskjold and O. Galteland,
\newblock{``Non-isothermal transport of multi-phase fluids in porous media. The entropy production,"}
\newblock{Frontiers in Physics} {\bf 6}, 126 (2018).

\bibitem{Gray_Miller_14}
W.G. Gray and C.T. Miller,
\newblock {``{I}ntroduction to the {T}hermodynamically {C}onstrained {A}veraging {T}heory for {P}orous {M}edium {S}ystems,"}
 (Springer, Z\"{u}rich, 2014).

\bibitem{Bear_Nitao_1995}
J. Bear and J.J., Nitao, 
\newblock{``On equilibrium and primary variables in transport in porous media."}
\newblock{Transp Porous Med} { \bf18,} 151--184 (1995).
%https://doi.org/10.1007/BF01064676

\bibitem{marle1981multiphase}
  {Marle, C.},
  \newblock{``Multiphase Flow in Porous Media"},
{Editions Technip} {(1981)}.

\bibitem{XU2006452}
J. Xu, S. Kjelstrup, D. Bedeaux, A. Røsjorde, L. Rekvig
\newblock{``Verification of Onsager's reciprocal relations for evaporation and condensation using non-equilibrium molecular dynamics,"}
\newblock{Journal of Colloid and Interface Science} {\bf 299}, 452--463 (2006).

\bibitem{Teller_1953}
H.A. Bethe and E. Teller,
\newblock{``Deviations from Thermal Equilibrium in Shock Waves,"}
(No. NP-4898; BRL-X-117). Engineering Research Institute, University of Michigan (1953). 
%https://doi.org/10.2172/4420349

\bibitem{Li_Garing_2020}
Y. Li, C. Garing and S.M. Benson, 
\newblock{``A continuum-scale representation of Ostwald ripening in heterogeneous porous media,"}
\newblock{Journal of Fluid Mechanics} {\bf 889} A14 (2020). 
%doi: 10.1017/jfm.2020.53.

\bibitem{Battiato_Tartakovsky_2011}
I. Battiato, D.M. Tartakovsky,
\newblock{``Applicability regimes for macroscopic models of reactive transport in porous media"},
\newblock{Journal of Contaminant Hydrology},
{\bf120–121}, 18--26 (2011).

\bibitem{Battiato_etal_2009}
I. Battiato, D.M. Tartakovsky, A.M. Tartakovsky, T. Scheibe,
\newblock{``On breakdown of macroscopic models of mixing-controlled heterogeneous reactions in porous media",}
\newblock{Advances in Water Resources}, {\bf32}, 1664--1673 (2009).
%https://doi.org/10.1016/j.adB. K.Jvwatres.2009.08.008.

\bibitem{Cushman_Ginn_1993}
J.H. Cushman and T.R.  Ginn, 
\newblock{``Nonlocal dispersion in media with continuously evolving scales of heterogeneity"}. 
\newblock{Transp Porous Med} {\bf 13}, 123--128 (1993).
%https://doi.org/10.1007/BF00613273

\bibitem{Cushman_OMalley_2015}
J. H. Cushman and D. O’Malley,
\newblock{``Fickian dispersion is anomalous"},
\newblock{Journal of Hydrology},
{\bf 531}, 161--167 (2015).
%https://doi.org/10.1016/j.jhydrol.2015.06.036.

\bibitem{Schluter_Berg_etal_2017}
 S. Schl\"{u}ter, Berg, S., Li, T., Vogel, H.‐J., and Wildenschild, D. 
\newblock{``Time scales of relaxation dynamics during transient conditions in two‐phase flow,"} 
\newblock{Water Resour. Res.}, {\bf 53}, 4709--4724 (2017).
%doi:10.1002/2016WR019815.

\bibitem{McClure_etal_2020}
McClure, J.E., Ramstad, T., Li, Z. Armstrong, R.T., Berg, S. 
\newblock{``Modeling Geometric State for Fluids in Porous Media: Evolution of the Euler Characteristic."} 
\newblock{Transp Porous Med } {\bf133}, 229--250 (2020). 
%https://doi.org/10.1007/s11242-020-01420-1

\bibitem{Orme_1997}
M. Orme,
\newblock{``Experiments on droplet collisions, bounce, coalescence and disruption,"}
\newblock{Progress in Energy and Combustion Sci.} {\bf 23}, 65--79 (1997).

\bibitem{Toro-Mendoza_2019}
J. Toro-Mendoza, O. Paredes-Altuve, M.A. Velasquez and D.N. Petsev
\newblock{``Reframing droplet coalescence: Identifying the distinctive dynamics of nanofilm evolution,"}
\newblock{Phys. Rev. Fluids} {\bf 4}, 093604 (2019).

\bibitem{Wu_2004}
M. Wu
\newblock{``Scaling law in liquid drop coalescence driven by surface tension,"}
\newblock{Physics of Fluids} {\bf 16}, L51 (2004).

\bibitem{Ristenpart_2006}
W. D. Ristenpart, P. M. McCalla, R. V. Roy, and H. A. Stone
\newblock{``Coalescence of Spreading Droplets on a Wettable Substrate,"}
\newblock{Phys. Rev. Lett.} {\bf 97}, 064501 (2006).

\bibitem{Dirk_2005}
Dirk G. A. L. Aarts, Henk N. W. Lekkerkerker, Hua Guo, G.H. Wegdam, D. Bonn
\newblock{``Hydrodynamics of Droplet Coalescence,"}
\newblock{Phys. Rev. Lett.} {\bf 95}, 164503 (2005).

\bibitem{Pauslon_NatureComm_2012}
J.D. Paulsen, R. Carmigniani, 
Anerudh Kannan, Justin C. Burton, and Sidney R. Nagel,
\newblock{``Coalescence of bubbles and drops in an outer fluid,"}
\newblock{Nat. Commun.} {\bf 5}, 3182 (2014).
%doi: 10.1038/ncomms4182 

\bibitem{Pauslon_PNAS_2012}
J.D. Paulsen, J.C. Burton, S.R. Nagel, S. Appathurai, M.T. Harris, O.A. Basaran,
\newblock{``The inexorable resistance of inertia determines the initial regime of drop coalescence,"}
\newblock{PNAS} {\bf 109}, 6857--6861 (2012).

\bibitem{Pauslon_PRL_2011}
J.D. Paulsen, J.C. Burton, S.R. Nagel,
%\newblock{``Viscous to Inertial Crossover in Liquid Drop Coalescence,"}
\newblock{Phys. Rev. Lett.} {\bf 106}, 114501 (2011).

\bibitem{Pauslon_PRL_2008}
S. C. Case and S. R. Nagel,
%\newblock{``Coalescence in low-viscosity liquids,"}
\newblock{Phys. Rev. Lett.} {\bf 100}, 084503 (2008).

\bibitem{Pak_etal_JPC_2018}
C.Y. Pak, W. Li, Y-L. Tse,
\newblock{``Free Energy and Dynamics of Water Droplet Coalescence,"}
\newblock{Journal of Physical Chemistry}, {\bf 122}, 22975--22984 (2018).
% provides molecular level view of thermodynamics
%https://pubs.acs.org/doi/10.1021/acs.jpcc.8b06507

\bibitem{Li_etal_nature_2018}
X. Li, H. Ren, W. Wu, H. Li, L. Wang, Y. He, J. Wang and Y. Zhou,
\newblock{``Wettability and Coalescence of Cu Droplets Subjected to Two-Wall Confinement,"}
\newblock{Scientific Reports}, {\bf 5}, 15190 (2015).

\bibitem{Vahabi_etal_science_2018}
H. Vahabi, W. Wang, J.M. Mabry and A.K. Kota,
\newblock{``Coalescence-induced jumping of droplets on superomniphobic surfaces with macrotexture,"}
\newblock{Science Advances}, {\bf 4}, 3488 (2018).
%https://www.pnas.org/content/111/21/7588

\bibitem{Pahlavan_2019}
A.A. Pahlavan, H.A. Stone, G.H. McKinley and R. Juanes
\newblock{``Restoring universality to the pinch-off of a bubble"},
\newblock{Proceedings of the National Academy of Sciences} {\bf116} (28), 13780--13784 (2019).

\bibitem{Perumananth_etal_PRL_2019}
S. Perumananth, M.K. Borg, M.V. Chubynsky, J.E. Sprittles, J.M. Reese,
\newblock{``Droplet Coalescence is Initiated by Thermal Motion,"}
\newblock{Physical Review Letters}, {\bf 122}, 104501 (2019).

\bibitem{Roof_1970}
J.G. Roof 
\newblock{``Snap-off of oil droplets in water-wet pores,"}
\newblock{SPEJ} {\bf 10}, 85--90 (1970)

\bibitem{Haines_1930}
W. B. Haines, 
\newblock{``Studies in the physical properties of soil. V. The hysteresis effect in capillary properties, and the modes of moisture distribution associated therewith"}, 
\newblock{J. Agric. Sci.}, {\bf 20}, 97--116, (1930).

\bibitem{Adamson_Gast_97}
A.W. Adamson and A.P. Gast,
\newblock {{``Physical Chemistry of Surfaces,"}}
 (Wiley, Hoboken, 1997).

\bibitem{Seth_Morrow_2007}
S. Seth and N.R Morrow,
\newblock{``Efficiency of the conversion of work of drainage to surface energy for sandstone and carbonate,"}
\newblock{SPE Reservoir Evaluation \& Engineering} {\bf 10}, 338–347 (2007).

\bibitem{Berg_Slotte_2020}
C.F. Berg, P.A Slotte and H.H. Khanamiri,
\newblock{Phys. Rev. E.}, {\bf 102}, 033113 (2020)

\bibitem{Gray_Dye_etal_15}
 W.G.Gray, A.L. Dye, J.E. McClure, L. J. Pyrak-Nolte and C.T.Miller,
\newblock {``On the dynamics and kinematics of two‐fluid‐phase flow in porous media,"}
\newblock { {Water Resour. Res.}}, {\bf 51}, 5365--5381 {(2015)}.

\bibitem{Kjelstrup_Bedeaux_08}
S. Kjelstrup and D. Bedeaux,
\newblock{``Non-Equilibrium Thermodynamics of Heterogeneous Systems,"}
(World Scientific, Hackensack, 2008)

\bibitem{Galteland_2019}
O. Galteland, D. Bedeaux, S. Kjelstrup and B. Hafskjold,
\newblock{``Pressures inside a nano-porous medium. The case of a single phase fluid,"}
\newblock{Frontiers in Physics} {\bf 7}, 60, (2019).

\bibitem{Mohanty_etal_1987}
K. K. Mohanty, H. T. Davis and L.E. Scriven, 
\newblock{``Physics of Oil Entrapment in Water-Wet Rock."} 
\newblock{Society of Petroleum Engineers} 
{\bf 2}, 113--128 (1987).
 %doi:10.2118/9406-PA
 
\bibitem{DiCarlo_2003}
D.A. DiCarlo, J.I.G Cidoncha and C. Hickey,
\newblock{``Acoustic measurements of pore-scale displacements."}
\newblock{Geophysical Research Letters}, {\bf 30}, 1901 (2003).
%doi:10.1029/2003GL017811, 
 
\bibitem{Aryana_2013}
S.A. Aryana and A.R. Kovscek, 
\newblock{``Nonequilibrium Effects and Multiphase Flow in Porous Media."}
\newblock{Transp. Porous Med.} {\bf 97}, 373--394 (2013). % https://doi.org/10.1007/s11242-013-0129-y

\bibitem{McClure_etal_PRL_submitted}
J.E. McClure, S. Berg and R.T. Armstrong, 
\newblock{``Multiscale fluctuations in non-equilibrium systems,"} 
\newblock{arXiv } {\bf TBD}, TBD (2020). 

\bibitem{Whitaker_1986}
 S. Whitaker,
\newblock{``Flow in porous media II: The governing equations for immiscible, two-phase flow."}
\newblock{Transp Porous Med}
{\bf 1}, 105--125 (1986). 
%https://doi.org/10.1007/BF00714688

\bibitem{Purswani2019}
P. Purswani,  M.S. Tawfik, Z.T. Karpyn, R.T. Johns
\newblock{``On the development of a relative permeability equation of state,"}
\newblock{Computational Geosciences},
{\bf 24}, 807--818 (2019).
%doi = {10.1007/s10596-019-9824-2}

\bibitem{Amaziane_Milisic_etal_12}
B. Amaziane, J.P. Milisic, M. Panfilov, and L. Pankratov,
\newblock {``Generalized nonequilibrium capillary relations for two-phase flow  through heterogeneous media,"}
\newblock { {Phys. Rev. E}}, {\bf85}, {016304} {(2012)}.

\bibitem{Lenormand_1988}
R. Lenormand, E. Touboul, C. Zarcone
\newblock{``Numerical models and experiments on immiscible displacements in porous media,"}
\newblock{J. Fluid Mech.} {\bf 189} 165--187, 1988

\bibitem{Rucker_2015}
M. R\"{u}cker, S. Berg, R. T. Armstrong, A. Georgiadis, H. Ott, A. Schwing, R. Neiteler, N. Brussee, A. Makurat, L. Leu, M. Wolf, F. Khan, F. Enzmann, M. Kersten 
\newblock{``From connected pathway flow to ganglion dynamics,"}
\newblock{Geophysical Research Letters} 42, 3888--3894, (2015).
 
\bibitem{McClure_Armstrong_etal_2018}
J.E. McClure, R.T. Armstrong, M.A. Berrill, S. Schl\"{u}ter, S. Berg, W.G. Gray, C.T. Miller,
\newblock{``Geometric state function for two-fluid flow in porous media,"}
\newblock{Phys. Rev. Fluids} {\bf 3} 084306 (2018).

%\bibitem{Xu_Louge_15}
%J. Xu, and M.Y. Louge,
%\newblock {Phys. Rev. E}, {\bf92}, 062405 (2015).

\bibitem{Schlueter2016b}
S. Schl\"{u}ter, S.  Berg, S. M. R\"{u}cker, R.T. Armstrong,
H.-J. Vogel, R. Hilfer, D. Wildenschild,
\newblock{``Pore‐scale displacement mechanisms as a source of hysteresis for two‐phase flow in porous media,"}
\newblock{Water Resour. Res.}, {\bf 52}, {2194--2205} {(2016)}

\bibitem{Land_68}
C.S. Land
\newblock{``Calculation of imbibition relative permeability for two- and three-phase flow from rock properties,"}
\newblock{SPEJ}, {\bf 8}, 149--156, (1968).

\bibitem{Lenhard_Parker_87}
R.J. Lenhard, J.C. Parker
\newblock{``A model for hysteretic constitutive relations governing multiphase flow,"}
\newblock{Water Resour. Res.}, {\bf 23}, 2187--2196, (1987).

\bibitem{Hilfer_06}
R. Hilfer,
\newblock {``Macroscopic capillarity and hysteresis for flow in porous media,"}
\newblock { {Phys. Rev. E}}, {\bf73}, {016307} {(2006)}.

\bibitem{Liu_Herring_etal_17}
Z. Liu, A. Herring, C. Arns, S. Berg, and R.T. Armstrong,
\newblock{``Pore-Scale Characterization of Two-Phase Flow Using Integral Geometry,"}
\newblock{Transp. Porous Media}, {\bf 1},  99--117 (2017).
%  publisher={Springer Netherlands}

\bibitem{Armstrong_McClure_etal_16}
R.T. Armstrong, J.E. McClure, M.A. Berrill, M. R\"{u}cker, S. Schl\"{u}ter, S. Berg,
\newblock{``Beyond Darcy's law: The role of phase topology and ganglion dynamics for two-fluid flow,"}
\newblock { {Phys. Rev. E}}, {\bf 94}, 043113 {(2016)}.

\bibitem{Herring_Harper_etal_13}
A.L. Herring, E.J. Harper, L. Andersson, A. Sheppard,
  B.K. Bay, D. Wildenschild,
\newblock{``Effect of fluid topology on residual nonwetting phase trapping: Implications for geologic co2 sequestration,"}
\newblock { Adv. Water Res.}, {\bf62}, 47--58, (2013).

\bibitem{McClure_Berrill_etal_16b}
J.E. McClure, M.A. Berrill, W.G. Gray, and C.T. Miller,
\newblock {``Influence of phase connectivity on the relationship among capillary pressure, fluid saturation, and interfacial area in two-fluid-phase porous medium systems,"}
\newblock { {Phys. Rev. E}}, {\bf 94}, 033102 {(2016)}.

\bibitem{Gray_Leijnse_1993}
W.G. Gray, A. Leijnse, 
\newblock {``{Mathematical tools for changing spatial scales in the
analysis of physical systems},"}
 (CRC Press, Boca Raton, 1993).


\bibitem{mcclure2020lbpm}
J.~E. McClure, Z. Li, M.A. Berrill, and T. Ramstad,
\newblock {``The LBPM software package for simulating multiphase flow on digital images of porous rocks,"} 
\newblock{Computational Geosciences}, 25, 871–895 (2021).

\bibitem{Veveakis_2015}
E. Veveakis, K. Regenauer-Lieb,
\newblock{``Review of extremum postulates,"}
\newblock{Current Opinion in Chemical Engineerging} {\bf 7}, 40--46 (2015)

\bibitem{Flekkoy_1999}
E.G. Flekk{\o}y, S.R. Pride,
\newblock{``Reciprocity and cross coupling of two-phase flow in porous media from Onsager theory"}
\newblock{Physical Review E} {\bf 60}, 4130 (1999)


\bibitem{Alpak_2019}
F. O. Alpak, I. Zacharoudiou, S. Berg, J. Dietderich, N. Saxena,
\newblock{``Direct simulation of pore-scale two-phase visco-capillary flow on large digital rock images using a phase-field lattice Boltzmann method on general-purpose graphics processing units."}
\newblock{Computational Geosciences}{\bf 23} 849--880 (2019).

\bibitem{Bak_1987}
 P. Bak,  C. Tang and K. Wiesenfeld,
 \newblock{``Self-organized criticality: an explanation of 1/f noise,"} 
 \newblock{Phys Rev Let.} {\bf 59} 381 (1987). 
 %doi:10.1103/PhysRevLett.59.381. 

\end{thebibliography}

\end{document}